\documentclass{Araya}

\usepackage[utf8]{inputenc} % allow utf-8 input
\usepackage[T1]{fontenc}    % use 8-bit T1 fonts
\usepackage{hyperref}       % hyperlinks
\usepackage{url}            % simple URL typesetting
\usepackage{booktabs}       % professional-quality tables
\usepackage{amsfonts}       % blackboard math symbols
\usepackage{nicefrac}       % compact symbols for 1/2, etc.
\usepackage{microtype}      % microtypography
\usepackage{fancyhdr}       % header
\usepackage{graphicx}       % graphics

\usepackage{amsmath}
\usepackage{amsthm}
\usepackage{mathrsfs}
\usepackage{mathtools}
\usepackage{stmaryrd}
\usepackage[capitalize]{cleveref}
\usepackage{tikz}
\usepackage{tikzit}
\usepackage{quiver}
\usepackage[dvipsnames,svgnames,table]{xcolor}

\newcommand{\supp}{\operatorname{supp}}
\newcommand{\KL}{D_{\mathrm{KL}}}
\DeclarePairedDelimiterX{\infdivx}[2]{(}{)}{%
  #1\;\delimsize\|\;#2%
}
\newcommand{\infdiv}{\KL \infdivx}

\renewcommand{\function}{g}               % generic function, f is already used below

\newcommand{\comp}{\mathop{\fatsemi}}

\newcommand{\Cat}{\mathbf{C}}
\renewcommand{\hom}{\mathrm{hom}}
\newcommand{\ob}{\mathrm{ob}}

\newcommand{\CatC}{\mathbf{C}}
\newcommand{\CatD}{\mathbf{D}}
\newcommand{\catset}{\mathbf{Set}}
\newcommand{\Coalg}{\mathbf{Coalg}}

\newcommand{\Func}{\mathcal{F}}
\newcommand{\idFunctor}{\mathsf{Id}}

\newcommand{\transitionDSInputs}{\beta}
\newcommand{\observationsDS}{\delta}

\newcommand{\type}{\mathcal{F}}
\newcommand{\dist}{p}                   % element of the set below
\newcommand{\Dist}{P}                   % set of probabilities distributions for the definition of the distribution functor
\newcommand{\Power}{\mathcal{P}}
\newcommand{\PowerFinite}{{\Power_\text{fin}}}
\newcommand{\homomorphism}{\phi}
\newcommand{\invhomomorphism}{\homomorphism^{-1}}

\newcommand{\Bisim}{B}

\newcommand{\statesFinal}{\omega}
\newcommand{\StatesFinal}{\Omega}
\newcommand{\transitionCoalgFinal}{\omega}
\newcommand{\beh}{\mathsf{beh}}

\newcommand{\states}{s}
\newcommand{\States}{S}
\newcommand{\statesPrime}{{s'}}
\newcommand\StatesPrime{{S'}}
\newcommand{\observations}{o}
\newcommand{\Observations}{O}

\newcommand{\inputs}{i}
\newcommand{\Inputs}{I}

\newcommand{\beliefsFiltering}{z}
\newcommand{\BeliefsFiltering}{Z}
\newcommand{\beliefsFilteringPrime}{{\beliefsFiltering'}}
\newcommand\BeliefsFilteringPrime{{\BeliefsFiltering'}}
\newcommand{\beliefsPredicting}{w}
\newcommand{\BeliefsPredicting}{W}
\newcommand{\beliefsPredictingPrime}{{\beliefsPredicting'}}
\newcommand\BeliefsPredictingPrime{{\BeliefsPredicting'}}
\newcommand{\beliefUpdateFiltering}{\tau_\BeliefsFiltering}

\newcommand{\historiesFiltering}{h}
\newcommand{\HistoriesFiltering}{H}

\newcommand{\transitionCoalg}{f}
\newcommand{\transitionCoalgPrime}{f'}
\newcommand{\transitionCoalgTr}{\mathsf{tr}}
\newcommand{\transitionCoalgOut}{\mathsf{out}}

\newcommand{\transitionBisim}{\gamma}                % bisimulation transition

\newcommand{\moore}{\mathsf{Moore}}

\newcommand{\pomdp}{{\mathsf{POMDP}}}

\newcommand{\mdp}{{\mathsf{MDP}}}

\newcommand{\powermoore}{{\PowerFinite\mathsf{Moore}}}

\title{A coalgebraic perspective on predictive processing
}

\author%
{\href{https://orcid.org/0000-0002-6086-4711}{\includegraphics[scale=0.06]{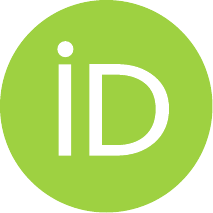}\hspace{1mm}Manuel Baltieri}~$^{1,2,}$\footnote[2]{Correspondence e-mail: \texttt{manuel\_baltieri@araya.org}},
\href{https://orcid.org/0000-0001-8790-2565}{\includegraphics[scale=0.06]{figures/orcid.pdf}\hspace{1mm}Filippo Torresan}~$^{1,2}$, \href{https://orcid.org/0000-0001-5225-0894}{\includegraphics[scale=0.06]{figures/orcid.pdf}\hspace{1mm}Tomoya Nakai}~$^{1}$\\
\vspace{1em} % Space between authors and afilliations
\normalfont{\small $^{1}$ Araya Inc., Tokyo, Japan}\\
\normalfont{\small $^{2}$ University of Sussex, Brighton, UK}\\
}

\begin{document}
\maketitle

\begin{abstract}
    Predictive processing and active inference posit that the brain is a system performing Bayesian inference on the environment.
    By virtue of this, a prominent interpretation of predictive processing states that the generative model (a POMDP) encoded by the brain synchronises with the generative process (another POMDP) representing the environment while trying to explain what hidden properties of the world generated its sensory input.
    In this view, the brain is thought to become a copy of the environment.
    This claim has however been disputed, stressing the fact that a structural copy, or isomorphism as it is at times invoked to be, is not an accurate description of this process since the environment is necessarily more complex than the brain, and what matters is not the capacity to exactly recapitulate the veridical causal structure of the world.
    In this work, we make parts of this counterargument formal by using ideas from the theory of coalgebras, an abstract mathematical framework for dynamical systems that brings together work from automata theory, concurrency theory, probabilistic processes and other fields.
    To do so, we cast generative model and process, in the form of POMDPs, as coalgebras, and use maps between them to describe a form of consistency that goes beyond mere structural similarity, giving the necessary mathematical background to describe how different processes can be seen as behaviourally, rather than structurally, equivalent, i.e. how they can be seen as emitting the same observations, and thus minimise prediction error, over time without strict assumptions about structural similarity.
    In particular, we will introduce three standard notions of equivalence from the literature on coalgebras, evaluating them in the context of predictive processing and identifying the one closest to claims made by proponents of this framework.
\end{abstract}

% keywords can be removed
\keywords{active inference, action-oriented models, coalgebras, bisimulations, behavioural equivalence}

%
%
%
% -----------------------------------------------------------------------------------------
% -----------------------------------------------------------------------------------------
% -----------------------------------------------------------------------------------------

\section{Introduction}

Predictive processing, and its more general formulation known as active inference, have been proposed as a general computational theory to account for the functions of the nervous system~\cite{fristonLearningInferenceBrain2003, fristonTheoryCorticalResponses2005}.
The proposal's key claim is that one can understand brain activity in its various forms and manifestations as resulting from the single imperative of minimising (variational) free energy~\cite{fristonPredictiveCodingFreeenergy2009}.
Thus, active inference promises to be a unified account of cognition and sentient behaviour, explaining in particular how key cognitive functions such as perception, action, and learning all emerge from a single principle, i.e. free energy minimisation~\cite{fristonFreeenergyPrincipleUnified2010, clarkWhateverNextPredictive2013, clarkSurfingUncertaintyPrediction2015, hohwyPredictiveMind2013, adamsPredictionsNotCommands2013, parrUncertaintyEpistemicsActive2017, torresanActiveInferenceActionunaware2025}.

In this view, the brain is described as a prediction machine~\cite{clarkRadicalPredictiveProcessing2015, clarkSurfingUncertaintyPrediction2015}, and every living organism is thought to be constantly trying to match or predict incoming sensory inputs produced by the environment, described as a generative process, that are relevant to itself.
If portions of the sensory data remain unaccounted for, then prediction errors ensue.
Variational inference is a general framework from probabilistic machine learning that, under certain assumptions, reduces to prediction error minimisation.
Variational free energy is a measure quantifying how much unexpected current sensory inputs are, with respect to the generative model and its updates, encoded by the nervous system (see the next section for a more formal treatment of these notions).
A generative model can be seen as an approximate, probabilistic representation of the surrounding environment encoded by an agent, describing how sensations and observations arise for a certain organism depending on context and behaviour (actions)~\cite{parrDiscreteContinuousBrain2018, parrGenerativeModelsActive2021, clarkDreamingWholeCat2012, fristonLearningInferenceBrain2003}.
By encoding updates consistent with an implicit hierarchical generative model, an agent's nervous system is thought to implement predictions, combined with a set of prior beliefs, about the most likely sensory inputs in a certain environment~\cite{kieferContentMisrepresentationHierarchical2018, kieferRepresentationPredictionError2019, fristonVariationalTreatmentDynamic2008}.

But what is the precise relation between the generative process and generative model?
In the literature, this is somewhat unclear: standard treatments of active inference and predictive processing often invoke a notion akin to structural similarity between the two, based on the idea that an agent ought to recapitulate the (statistical, or at times causal) structure of the environment~\cite{hohwyPredictiveMind2013, gladziejewskiPredictiveCodingRepresentationalism2016, ramsteadTaleTwoDensities2019, kieferContentMisrepresentationHierarchical2018, kieferRepresentationPredictionError2019, ramsteadNeuralPhenotypicRepresentation2021}, and that the two must somehow synchronise~\cite{fristonLifeWeKnow2013, senguptaSentientSelfOrganizationMinimal2017, fristonFreeEnergyPrinciple2019, palaciosEmergenceSynchronyNetworks2019, parr2020markov, dacostaActiveInferenceDiscrete2020, parrModulesMeanFields2020, fristonStochasticChaosMarkov2021, parrActiveInferenceFree2022, fristonFreeEnergyPrinciple2023}.
Other works have on the other hand argued that structural similarity is not necessary, in the sense that one can formulate \emph{action-oriented} generative models that do not capture the structural richness of their respective generative processes~\cite{clarkRadicalPredictiveProcessing2015, clarkSurfingUncertaintyPrediction2015, baltieriActiveInferenceImplementation2017, baltieriGenerativeModelsParsimonious2019, baltieriPIDControlProcess2019, tschantzLearningActionorientedModels2020, mannellaActiveInferenceWhiskers2021} (see also~\cite{dacostaActiveInferenceDiscrete2020}, stating that ``[h]idden and external states may or may not be isomorphic [\dots]'').

To better explain the perspective invoked by the latter, largely based on informal accounts~\cite{clarkRadicalPredictiveProcessing2015, clarkSurfingUncertaintyPrediction2015} and simulation work~\cite{baltieriActiveInferenceImplementation2017, baltieriGenerativeModelsParsimonious2019, baltieriPIDControlProcess2019, tschantzLearningActionorientedModels2020, mannellaActiveInferenceWhiskers2021}, we reformulate the intuition behind it using the theory of coalgebras, treating generative process and models as coalgebras, and exploring their relation in terms of specific maps between them.
Coalgebras are a standard tool used in theoretical computer science and mathematics (especially category theory)~\cite{ruttenUniversalCoalgebraTheory2000, jacobsIntroductionCoalgebraMathematics2017} for the treatment of general dynamical systems.
Their structural formulation puts an emphasis on objects (dynamical systems) and maps between them (homomorphisms) that must satisfy certain conditions.
Thanks to their abstract definitions, their expressive power is quite broad and will allow us to easily bring out examples of growing difficulty for different transition types (with inputs, outputs or both, terminal states, etc.) and branching in dynamical systems (deterministic, possibilistic and probabilistic), focusing in the end on the particular implementations relevant for this work: POMDPs in active inference, and behavioural equivalence between them.

In~\cref{sec:activeInference}, we provide a structural overview of active inference with a focus on its core working parts and its applications to perception, i.e. predictive processing. 
\cref{sec:categoriesCoalgebras} offers a short, self-contained introduction to coalgebras and their maps, with an emphasis on definitions of bisimulation (equivalence), kernel bisimulation and behavioural equivalence and a few standard examples of their uses.
In~\cref{sec:PPcoalgebras}, we bring these two parts together, with POMDPs in active inference formulated as coalgebras, and their relation as a kind of consistency between coalgebras using the aforementioned definitions.

%
%
%
% -----------------------------------------------------------------------------------------
% -----------------------------------------------------------------------------------------
% -----------------------------------------------------------------------------------------

\section{Predictive processing, an overview}
\label{sec:activeInference}
While a full treatment of active inference remains outside the scope of the present manuscript, in this section we introduce the main motivations behind this framework, which will allow us to formally discuss its main structural components, showing connections between active inference and the theory of coalgebras.
For some technical treatments and reviews, see e.g.~\cite{bogaczTutorialFreeenergyFramework2017, buckleyFreeEnergyPrinciple2017, fristonActiveInferenceProcess2017, biehlExpandingActiveInference2018, dacostaActiveInferenceDiscrete2020, fristonSophisticatedInference2021, parrActiveInferenceFree2022}.

%
%
% ---------------------------------------------------
% ---------------------------------------------------

\subsection{The structure of active inference problems}
\label{sec:activeInferenceStructure}
\begin{figure}
    \centering
    \includegraphics[width=0.7\linewidth]{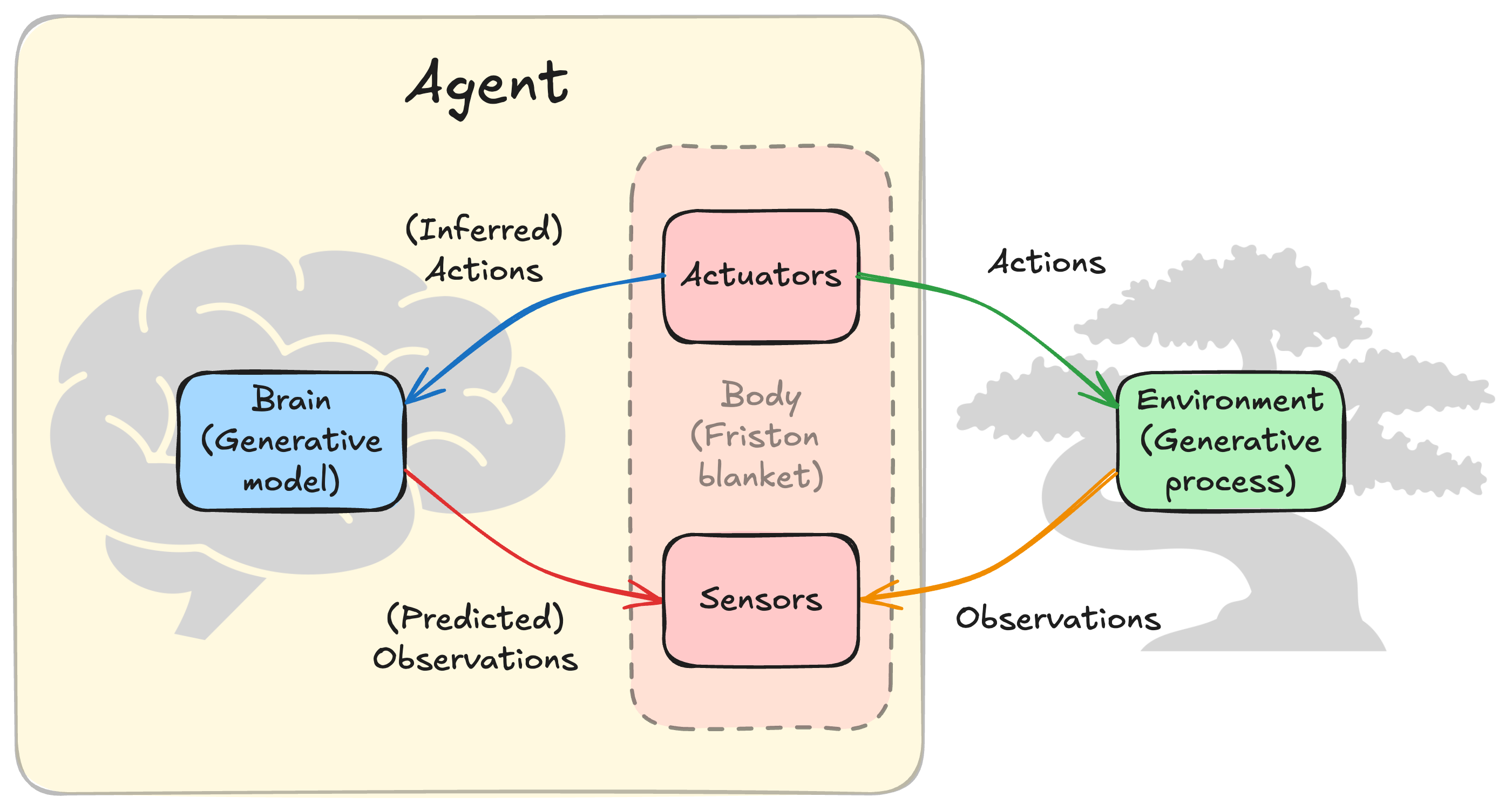}
    \caption{\textbf{Active inference setup}. Brain-body-environment factorisation of an agent.}
    \label{fig:brainBodyEnv}
\end{figure}

In a standard active inference setup, we have an agent actively interacting with an environment.
The agent can be described as a system factored into two components: a brain and body, a setup typical of embodied approaches to biology and cognitive science~\cite{beerDynamicalApproachesCognitive2000, beerDynamicsBrainBody2008}, see~\cref{fig:brainBodyEnv}.
Note that in principle, these need not be a ``brain'' and a ``body'' in a strict sense; one can for instance imagine an E. Coli's signalling pathway~\cite{alonRobustnessBacterialChemotaxis1999a, andrewsOptimalNoiseFiltering2006} to play the role of a brain in this setup.

%
% ----------------------------------

\paragraph{Environment}
Focusing on the discrete-time treatments ~\cite{fristonActiveInferenceProcess2017, biehlExpandingActiveInference2018, sajidActiveInferenceDemystified2021, dacostaActiveInferenceDiscrete2020, fristonSophisticatedInference2021, parrActiveInferenceFree2022}, the relevant part of the environment with which an agent interacts, and which generates observations coming through its sensors, is referred to as \emph{generative process}.
A generative process is usually formulated as a partially observable Markov decision process~\cite{putermanMarkovDecisionProcesses2014, kaelblingPlanningActingPartially1998, suttonReinforcementLearningIntroduction2018}, and represents the ground truth for an agent:
\begin{definition}[Partially observable Markov decision process (POMDP)]
    \label{def:POMDP}
    A partially observable Markov decision process is a tuple 
    $(\States, \Actions, \TransitionMap, \Observations, \ObservationMap)$, where:
    \begin{itemize}
        \item $\States$ is the state space,
        \item $\Actions$ is the action space,
        \item $\TransitionMap : \States \times \Actions \to \Prob(\States)$ is the transitions dynamics, where $\Prob(\States)$ is the set of distributions over $\States$ with finite support such that for a given state $\states_{t}$ and $\actions_{t}$, $\TransitionMap(\states_{t}, \actions_{t})$ gives a probability distribution of states an agent can transition to from state $\states_{t}$ while taking action $\actions_{t}$, often written as $\prob(\states_{t+1} \mid \states_{t}, \actions_{t})$,
        \item $\Observations$ is the observation space,
        \item $\ObservationMap: \States \to \Prob(\Observations)$ is the observation map where $\Prob(\Observations)$ is the set of distributions over $\Observations$ with finite support such that for a given state $\states_{t}$, $\ObservationMap(\states_{t})$ gives a probability distribution on observations $\prob(\observations_{t})$.
    \end{itemize}
\end{definition}

Note that this corresponds to a ``POMDP without rewards''.
A more general definition of POMDP includes in fact:
\begin{itemize}
    \item $\discount \in [0,1)$ is a discount factor,
    \item $\reward: \States \times \Actions \to \Reals$, a map giving a reward every time a transition is taken.
\end{itemize}

This is because generative processes in active inference do not include a reward function (and thus no discount factor either).
One could say that rewards are effectively folded into $\Observations$, seen as observation-reward pairs $\Observations \coloneqq \Observations' \times \Reals$. However, it is standard practice in active inference to avoid reward functions, in favour of priors on (desired) outcomes/observations~\cite{fristonReinforcementLearningActive2009, fristonWhatOptimalMotor2011, fristonFreeEnergyValue2012, millidgeRelationshipActiveInference2020, sajidActiveInferenceDemystified2021, parrActiveInferenceFree2022}. 
See also~\cite{dacostaRewardMaximizationDiscrete2023} where, under the assumption that an agent has access to preferences that encode exactly rewarding states, a formulation compatible with standard (PO)MDPs can be derived.

%
% ----------------------------------

\paragraph{Body}
The body is often represented as an interface, equivalent to channels that couple the agent to the environment, in terms of sensors and actuators, forming what is usually referred to as an action-perception loop or sensorimotor loop, see for instance~\cite{vonuexkullStrollWorldsAnimals1992, ayCausalStructureSensorimotor2014}.
In the active inference literature, these components constitute the so-called ``Markov blanket'' of an agent~\cite{fristonParcelsParticlesMarkov2021, parrActiveInferenceFree2022, fristonFreeEnergyPrinciple2023}, depicted in~\cref{fig:brainBodyEnv} as a ``Friston blanket'' instead, following arguments found in~\cite{bruinebergEmperorsNewMarkov2022} (see also~\cite{biehlTechnicalCritiqueParts2021, rosasCausalBlanketsTheory2020, aguileraHowParticularPhysics2022, virgoEmbracingSensorimotorHistory2022}).

%
% ----------------------------------

\paragraph{Brain}
The role of the brain in this framework is to perform inference about unknown properties (states, parameters) of the environment. It acts as a prediction machine, as is often highlighted in the field of cognitive (neuro)science, generating the same observations it receives from the environment.
This is achieved by describing brain states as parametrising distributions of states and parameters of a \emph{generative model}, with their dynamics being consistent with belief updates in a Bayesian framework.
Generative models are a standard component of (Bayesian) machine learning setups~\cite{bishopPatternRecognitionMachine2006, murphyMachineLearningProbabilistic2014}, consisting of a joint probability distribution of observations emitted by the environment, and hidden variables/parameters that generated them.
We note that, under this interpretation, the brain does not contain, have, or even is a generative model.
Instead, brain states and their dynamics implement a scheme consistent with update equations that could be \emph{interpreted}~\cite{virgoInterpretingDynamicalSystems2021, biehlInterpretingSystemsSolving2022} as the brain implicitly having such a model~\cite{baltieriBayesianInterpretationInternal2025, virgoGoodRegulatorTheorem2025}.

In this context, a generative model consists of another POMDP~\cite{fristonActiveInferenceProcess2017, sajidActiveInferenceDemystified2021, dacostaActiveInferenceDiscrete2020, fristonSophisticatedInference2021, parrActiveInferenceFree2022}, whose goal is to approximate (and in an ideal, perfect model scenario, to match) in some sense the generative process, another POMDP.
This particular POMDP ought to have the same interface as the generative process POMDP.
According to active inference, an agent's goal is for its generative model to have the same inputs, i.e. actions generated for the environment, either by having a copy of them (efference copy) or by inferring them (as in more standard active inference setups, see~\cite{torresanActiveInferenceActionunaware2025}  and~\cref{fig:brainBodyEnv}), and the same outputs, i.e. predictions of observations that minimise variational free energy/prediction error (see~\cref{fig:brainBodyEnv}).

This setup is consistent with model-based approaches to reinforcement learning~\cite{suttonReinforcementLearningIntroduction2018} since the agent operates on the assumptions induced by the generative model.
Computationally, however, the goals and algorithmic implementations of these two approaches are usually quite different, although more recently converging to similar ideas~\cite{tschantzScalingActiveInference2020, tschantzReinforcementLearningActive2020, millidgeRelationshipActiveInference2020, imohiosenActiveInferenceControl2020, dacostaRewardMaximizationDiscrete2023, malekzadehActiveInferenceReinforcement2024}, mainly the free energy (and expected free energy) minimisation.
%
%
% ---------------------------------------------------
% ---------------------------------------------------

\subsection{Free energy minimisation and predictive processing}
One of the main goals of an active inference agent is to infer the most likely environment's configuration given a sequence of observations.
The task is formalised in terms of approximate Bayesian inference via variational Bayes.
To see the core idea, consider the application of Bayes' rule to a POMDP within the active inference framework:
\begin{equation}\label{eqn:aif-bayes-pomdp}
    \condprob{\seqv{\statevar}{1}{\ntime}, \policy, \obsmap, \transmap}{\seqv{\obsvar}{1}{\ntime}} = \frac{\condprob{\seqv{\obsvar}{1}{\ntime}}{\seqv{\statevar}{1}{\ntime}, \policy, \obsmap, \transmap} \Prob(\seqv{\statevar}{1}{\ntime}, \policy, \obsmap, \transmap)}{\Prob(\seqv{\obsvar}{1}{\ntime})},
\end{equation}
where $\policy$ is a policy, while $\obsmap, \transmap$ are parameters for observation and transition maps, respectively.
In general, this cannot be solved analytically.
Therefore, in active inference, the update of the agent's probabilistic beliefs is performed via an optimisation procedure involving a quantity called \emph{variational free energy}.
Variational free energy is defined in terms of a probability distribution known as the variational or approximate posterior, \(\varprob{\statevar_{1:\ntime}, \policy, \obsmap, \transmap}\), which approximates the true posterior in~\eqref{eqn:aif-bayes-pomdp}, giving the following: 
\begin{align}
    \label{eqn:aif-fe-pomdp}
    \fe \bigl[ \varprob{\statevar_{1:\ntime}, \policy, \obsmap, \transmap} \bigr] \coloneqq \mathbb{E}_{Q} \Bigl[\log \varprob{\statevar_{1:\ntime}, \policy, \obsmap, \transmap} - \log \Prob( \obsvar_{1:\ntime}, \statevar_{1:\ntime}, \policy, \obsmap, \transmap) \Bigr].
\end{align}
It can be shown that minimising variational free energy with respect to the parameters of those approximate posteriors is equivalent to performing approximate Bayesian inference. As a result, the variational posteriors become more aligned with the exact posteriors given the generative model that we were aiming to compute in the first place. 
This is expressed more concisely by the following:
\begin{align}
    \label{eqn:aif-kl-fe}
    \infdiv[\Big]{\varprob{\States_{1:\ntime}, \policy, \obsmap, \transmap}}{\condprob{\States_{1:\ntime}, \policy, \obsmap, \transmap}{\obsvar_{1:\ntime}}} = \fe \bigl[ \varprob{\States_{1:\ntime}, \policy, \obsmap, \transmap} \bigr] + \log \Prob(\obsvar_{1:\ntime}),
\end{align}
showing that minimising variational free is equivalent to minimising the KL divergence between the variational posterior and the exact (analytical) posterior, at least up to another term: the negative surprisal, $\log \Prob(\obsvar_{1:\ntime})$.
We note that the surprisal $- \log \Prob(\obsvar_{1:\ntime})$ can play different roles depending on the setup. In (machine) learning and perception, it is assumed to be constant, given by a fixed set of observations $\obsvar_{1:\ntime}$ that do not change over time.
However, when action is introduced, combining perception and action as in active inference, the surprisal itself can change by selectively sampling observations $\obsvar_{1:\ntime}$ through the choice of particular action policies.
In this work, we will focus on predictive processing, which primarily involves perception and learning processes in which the surprisal remains constant over time~\cite{clarkWhateverNextPredictive2013, clarkRadicalPredictiveProcessing2015, keller2018predictive, hohwyNewDirectionsPredictive2020}, and leave action to future work.

In contrast to the full expression of~\eqref{eqn:aif-bayes-pomdp}, free energy is a quantity that can be evaluated, given the specification of a generative model and the variational posterior, thus allowing an agent to use collected data/observations to update its probabilistic beliefs.
\Cref{eqn:aif-kl-fe} highlights an important aspect of this variational formulation: the starting point is the minimisation of the KL divergence between the approximate and real posteriors, which is achieved via minimising free energy under the assumption that the surprisal is assumed is fixed.
However, this does not yet specify what kind of relation ought to be in place between the (implicit) generative model encoded by the brain, and the generative process representing the ground truth of the environment.

In particular, minimising the free energy in~\cref{eqn:aif-fe-pomdp} only implies that the approximate posterior ought to be as close as possible to the true posterior obtained from the generative model given observations $\obsvar_{1:\ntime}$.
How do we ensure then that this process of approximate Bayesian inference is, in some sense, \emph{veridical} with respect to the underlying generative process, i.e. that a generative model is appropriate for a generative process?
In other words, how do we know that an agent minimising variational free energy will somehow be successful in a given the environment?

To answer this, we start with a simple observation: if a generative model is in some meaningful way \emph{wrong}, then free energy cannot be minimised to a satisfactory level.
This means that repeated attempts to implement processes of perception, learning, planning, and action selection, combined in a sensorimotor loop that aims to minimise free energy across time scales, are not sufficient for an agent to fulfil its preferences.
Active inference thus posits that further strategies should be in place in such cases, including, for instance, model selection~\cite{stephanBayesianModelSelection2009} and structure learning~\cite{smithActiveInferenceApproach2020}, so that a better generative model can be obtained.
The former refers to selecting a generative model from a pre-defined space of possible models that better match the requirements of a ``good'' generative model, one that allows an agent to fulfil its goals.
The latter involves combining expansion and reduction processes, which respectively create and remove variables in a generative model until it can more correctly be used to achieve a certain goal.

While the technical details are not particularly relevant in this work, the key message is that a generative model is only as good as the performance it enables an agent to achieve in the pursuit of its preferences.
In other words, the generative model must be \emph{close enough} to the generative process to capture the relevant aspects of the world that affect the agent's ability to realise its preferences.
Only then does it make sense to speak of minimising free energy for perception, learning, planning, and action selection.
Next we examine how the predictive processing literature has interpreted this idea of \emph{close enough} in different ways.

%
%
% -----------------------------------------------------------------------------------------
% -----------------------------------------------------------------------------------------

\subsection{Synchronisation of generative model and generative process?}
\label{sec:synchronisation}
There is some consensus in the active inference and predictive processing literature on the idea that the generative model of the agent recapitulates the (statistical, or at times causal) \emph{structure} of the generative process of the environment it refers to~\cite{hohwyPredictiveMind2013, gladziejewskiPredictiveCodingRepresentationalism2016, ramsteadTaleTwoDensities2019, dacostaActiveInferenceDiscrete2020, parrInferentialDynamics2022}.
However, the exact meaning of this statement, and the extent to which this ought to hold, are not entirely clear~\cite{clarkEmbodiedPrediction2015, baltieriActiveInferenceImplementation2017, tschantzLearningActionorientedModels2020, facchinStructuralRepresentationsNot2021, bruinebergEmperorsNewMarkov2022, aguileraHowParticularPhysics2022}, partly due to the ambiguity concerning what constitutes \emph{causal} structure (see, e.g.~\cite{torresanDisentangledRepresentationsCausal2024}), and partly because of the distinction between \emph{internal} and \emph{external} states~\cite{bruinebergEmperorsNewMarkov2022, bruinebergEmperorNakedReplies2022}.
In an attempt to formalise this idea, different works~\cite{fristonLifeWeKnow2013, senguptaSentientSelfOrganizationMinimal2017, fristonFreeEnergyPrinciple2019, palaciosEmergenceSynchronyNetworks2019, parr2020markov, dacostaActiveInferenceDiscrete2020, parrModulesMeanFields2020, fristonStochasticChaosMarkov2021, parrActiveInferenceFree2022, fristonFreeEnergyPrinciple2023} have argued that, under certain assumptions, mainly the existence of a synchronisation map~\cite{dacostaBayesianMechanicsStationary2021}, free energy minimisation entails a \emph{synchronisation} between an agent's internal states and external environment states. 
This synchronisation can be regarded as a kind of mirroring between internal and external states.

On the other hand, several works have argued~\cite{clarkRadicalPredictiveProcessing2015, clarkSurfingUncertaintyPrediction2015} and demonstrated~\cite{baltieriActiveInferenceImplementation2017, baltieriGenerativeModelsParsimonious2019, baltieriPIDControlProcess2019, tschantzLearningActionorientedModels2020, mannellaActiveInferenceWhiskers2021} that this structural synchronisation need not, in fact, hold. 
One can have generative models, often called \emph{action-oriented} in this area of research, that do not recapitulate the structural richness of their respective generative processes.
This is also now largely accepted by standard treatments of active inference, e.g.~\cite{dacostaActiveInferenceDiscrete2020}, stating that ``[h]idden and external states may or may not be isomorphic [\dots]'' and that ``an agent uses its internal states to represent hidden states that may or may not exist in the external world''.

To reconcile these seemingly divergent views, namely, that 1) a generative model and its respective generative process must synchronise, and that 2) the structure of the generative model can differ from that of the generative process, we take a perspective dual to the structural one: a perspective that puts \emph{behaviour} first.
The term ``behaviour'' tends to assume different connotations in different fields.
In psychology, for instance, it is often associated with behavioural research, focusing on the observable behaviour of a subject, its conditioning, and its interactions with the environment, in contrast to cognitive approaches that emphasise internal processes such as cognition and emotion.
In this work, we do not engage with this kind of debate, instead, we operationalise observable behaviour as outputs of a system over time.

This operationalisation stems from the well-known duality between structure and behaviour in theoretical computer science and mathematics~\cite{ruttenUniversalCoalgebraTheory2000, jacobsIntroductionCoalgebraMathematics2017}, where algebras are taken as a language best suited to describe structure, while \emph{co}algebras as a language for the behaviour of systems.
While this is by no means the only way to conceptualise systems and behaviours, it is a convenient approach that clearly highlights how to build a behavioural understanding of a system, starting from structural (i.e. algebraic) notions and adapting them (reversing arrows) by categorical duality~\cite{maclaneCategoriesWorkingMathematician1978}.
In the next section, we will provide a brief overview of coalgebras, and their role as a general language for state-based systems.
This includes both \emph{transition type} (dynamics and their possible effects such as termination, outputs, and inputs) and \emph{branching} (e.g. deterministic, probabilistic, or possibilistic systems) within the same framework in a convenient and formal way~\cite{hasuoGenericTraceSemantics2007}.

%
%
%
% -----------------------------------------------------------------------------------------
% -----------------------------------------------------------------------------------------
% -----------------------------------------------------------------------------------------

\section{Categories, coalgebras and bisimulations}
\label{sec:categoriesCoalgebras}
In this section we provide a brief overview of categorical ideas that lead to an abstract treatment of dynamical systems and maps between them.
This framework will later allow us to apply existing notions of behavioural consistency to processes of various kind, focusing in our particular setup on POMDPs.
Throughout this work, whenever we refer to a ``system'' we do so under the assumption that a system is a coalgebra.
While there are other doctrines and theories of systems based on various other definitions, see for instance~\cite{myersCategoricalSystemsTheory2021, capucciNotesCategoricalSystems2024, libkindDoubleOperadicTheory2025}, these are beyond the scope of the current work and will not be covered here.

%
%
% -----------------------------------------------------------------------------------------
% -----------------------------------------------------------------------------------------

\subsection{Categories and functors between them}
The first crucial idea involves the definition of category, an abstract collection of objects, and maps between them that are required to satisfy certain rules~\cite{maclaneCategoriesWorkingMathematician1978, riehlCategoryTheoryContext2017}.
\begin{definition}[Category]
    A category $\Cat$ consists of:
    \begin{itemize}
        \item a class of objects, $\ob(\Cat)$, e.g. $A, B, C, \dots$,
        \item a class of maps, arrows or morphisms, $\hom(\Cat)$, between objects (sources and targets), e.g. $f, g, h, \dots$, usually represented as $f: A \to B$, $g: B \to C$, $h: C \to D, \dots$,
        \item for each object of $\ob(\Cat)$, an identity morphism $\id_A: A \to A$,
        \item a binary operation $\comp$ representing the composition of morphisms, e.g. given the morphisms $f, g$ above, $f \comp g = A \to C$, that satisfies the following\footnote{Note that $\circ$ is also often use for composition, with $f \comp g = g \circ f$, we decided however to adopt $\comp$ as we think it helps following the order of a composition.},
        \begin{itemize}
            \item associativity, given $f, g, h$ above, $f \comp (g \comp h) = (f \comp g) \comp h$,
            \item left and right unit laws, for every pair of objects $A, B$ and morphism $f : A \to B$, $\id_A \comp f = f = f \comp \id_B$.
        \end{itemize}
    \end{itemize}
\end{definition}

Of particular interest for this work then, as well as for several other results in the field of category theory, is the concept of functor, a ``categorification'' (the process of replacing sets with categories to generalise set-theoretic definitions) of the standard notion of \emph{function} between sets.
\begin{definition}[Functor]
    Let $\CatC$ and $\CatD$ be categories. A (covariant) functor $\Func$ from $\CatC$ to $\CatD$ is a mapping that
    \begin{itemize}
        \item associates each object $X$ in $\CatC$ to an object $\Func(X)$ in $\CatD$,
        \item associates each morphism $f: X \to Y$ in $\CatC$ to a morphism $\Func(f): \Func(X) \to \Func(Y)$ in $\CatD$ such that the following two conditions hold:
        \begin{itemize}
            \item $\Func (\id_X) = \id_{\Func(X)}$ for every object $X$ in $\CatC$ (functors must preserve identity),
            \item $\Func(f \comp g) = \Func(f) \comp \Func(f)$ for all morphisms $f: X \to Y$ and $g: Y \to Z$ in $\CatC$ (functors must preserve composition of morphisms).
        \end{itemize}
    \end{itemize}
\end{definition}

For the theory of coalgebras, which we briefly overview next, we need to consider a special case of functor, called an \emph{endofunctor}, i.e. a functor from a category $\CatC$ to itself.

%
%
% ---------------------------------------------------
% ---------------------------------------------------

\subsection{Processes as coalgebras}
\label{sec:coalgebras}
Coalgebras, a construction from category theory related to algebras (which are their dual in a categorical sense~\cite{maclaneCategoriesWorkingMathematician1978}), have in recent years become a popular approach for the study of general dynamical systems and automata~\cite{ruttenUniversalCoalgebraTheory2000, jacobsIntroductionCoalgebraMathematics2017}.
Technically, given a category $\CatC$ and an endofunctor $\Func : \CatC \to \CatC$, we define an \mbox{$\type$-coalgebra} (or simply coalgebra, when $\type$ is understood) as an object $\States$ of $\CatC$ together with a map $\transitionCoalg^\States : \States \to \type(\States)$, represented as $(\States, \transitionCoalg^\States)$, where $\type$ is the type or signature of the coalgebra, $\States$ is the carrier, and $\transitionCoalg^\States$ is the (transition) structure map of the coalgebra.
Given a category $\CatC$ and an endofunctor $\Func$, we can build a category of coalgebras with \mbox{$\type$-coalgebras} of the form $(\States, \transitionCoalg^\States)$ as objects, and morphisms between them as maps.

\begin{definition}[Category of $\type$-coalgebras]
    \label{def:coalgebrasCat}
    $\Coalg_\CatC(\type)$ is the category of $\type$-coalgebras with objects \mbox{$\type$-coalgebras} and maps called $\type$-homomorphisms, coalgebra homomorphisms or simply homomorphism when the context allows it.
    Given objects $(\States, \transitionCoalg^\States), (\StatesPrime, \transitionCoalg^\StatesPrime)$ in $\Coalg_\CatC(\type)$, an $\type$-homomorphism $\homomorphism$ is a map that makes the following diagram commute:
    \begin{equation}
        % https://q.uiver.app/#q=WzAsNCxbMCwwLCJTIl0sWzAsMiwiRihTKSJdLFsyLDAsIlQiXSxbMiwyLCJGKFQpIl0sWzAsMSwiZiIsMl0sWzAsMiwiXFxwaGkiXSxbMiwzLCJnIl0sWzEsMywiRiAoXFxwaGkpIiwyXV0=
        \begin{tikzcd}
            \States && \StatesPrime \\
            \\
            {\type(\States)} && {\type(\StatesPrime)}
            \arrow["\transitionCoalg^\States"', from=1-1, to=3-1]
            \arrow["\homomorphism", from=1-1, to=1-3]
            \arrow["\transitionCoalg^\StatesPrime", from=1-3, to=3-3]
            \arrow["{\type (\homomorphism)}"', from=3-1, to=3-3]
        \end{tikzcd}
    \end{equation}
    i.e. such that $\transitionCoalg^\States \comp \type(\homomorphism) = \homomorphism \comp \transitionCoalg^\StatesPrime$.
    The identity is given by the trivial structure map $\States \xrightarrow{\id_\States} \States$ between the underlying sets. Composition is defined by placing commuting squares side by side, and associativity is established by verifying that the order in which they commute is not relevant.
\end{definition}

In this work, we focus exclusively on coalgebras where the base category is $\CatC = \catset$, the category whose objects are sets and whose morphisms are functions.
Accordingly, rather than using the somewhat bloated $\Coalg_\catset(\type)$, we will simplify the notation to $\Coalg(\type)$ for the category of coalgebras for the endofunctor $\type$ on the category $\catset$. 
As our treatment largely relies on pre-existing knowledge and intuitions about sets and functions, most other technical details necessary for a full categorical account of coalgebras will be skipped.

To develop a more intuitive understanding of coalgebras for the purpose of this work, i.e. discrete dynamical systems, we next introduce a few standard examples and definitions presented within this framework. 
We initially follow introductory treatments at first~\cite{ruttenUniversalCoalgebraTheory2000, ruttenMethodCoalgebraExercises2019}, and later refer to $\cite{silvaGeneralizingPowersetConstruction2010}$ for an example specifically relevant to the present work (i.e. involving probabilities).

%
% ----------------------------------

\subsubsection{Deterministic systems as coalgebras}
Deterministic systems constitute the simplest class of dynamical systems.
While these systems have a trivial branching (i.e. only one possible next state), their transition types can follow different rules. Here we consider both closed and open systems as illustrative examples of how different interfaces can be represented.
\begin{example}[Category of closed systems as coalgebras]
    \label{def:ClosedSystems}
    The category of closed systems as coalgebras $\Coalg(\idFunctor)$, with coalgebra type given by the identity functor $\idFunctor : \catset \to \catset$, has
    \begin{itemize}
        \item as objects, coalgebras of the form $(\States, \transitionCoalg^\States : \States \to \States)$, and
        \item as morphisms between two coalgebras $(\States, \transitionCoalg^\States)$ and $(\StatesPrime, \transitionCoalg^\StatesPrime)$, homomorphisms in the form of functions $\homomorphism: \States \to \StatesPrime$ that make the following diagram commute 
        \begin{equation}
            % https://q.uiver.app/#q=WzAsNCxbMCwwLCJTIl0sWzAsMiwiUyJdLFsyLDAsIlQiXSxbMiwyLCJUIl0sWzAsMSwiZiIsMl0sWzAsMiwiXFxwaGkiXSxbMiwzLCJnIl0sWzEsMywiXFxwaGkiLDJdXQ==
            \begin{tikzcd}
                \States && \StatesPrime \\
                \\
                \States && \StatesPrime
                \arrow["\transitionCoalg^\States"', from=1-1, to=3-1]
                \arrow["\homomorphism", from=1-1, to=1-3]
                \arrow["\transitionCoalg^\StatesPrime", from=1-3, to=3-3]
                \arrow["\homomorphism"', from=3-1, to=3-3]
            \end{tikzcd}
        \end{equation}
    \end{itemize}
\end{example}

Note that ``closed'' here refer to the fact that these systems have no inputs/outputs, that is, they are autonomous (no inputs) and have no observable outputs, and thus cannot communicate with the outside world (i.e. their interfaces are trivial). 
This, however, does not prevent us from describing maps that track consistent updates between different closed systems, maps that preserve transitions, i.e. ($\type$-)coalgebra homomorphisms.

On the other hand, Moore machines are classical architectures in the automata theory literature, corresponding to systems with inputs and outputs, or open discrete-time dynamical systems. 
Formally, a Moore machine is represented as a quintuple with states, inputs, outputs, transition, and output functions $(\States, \Inputs, \Observations, \transitionDSInputs: \Inputs \times \States \to \States, \observationsDS: \States \to \Observations)$.
\begin{example}[Category of Moore machines as coalgebras]
    \label{ex:MooreMachines}
    The category of Moore machines as coalgebras $\Coalg(\moore)$, with coalgebra type given by the functor $\moore : \catset \to \catset$ such that $\moore(\States) = \Observations \times \States^\Inputs$, has 
    \begin{itemize}
        \item as objects, coalgebras of the form $(\States, \transitionCoalg_\moore^\States : \States \to \Observations \times \States^\Inputs)$, where $\States^\Inputs$ is the set of all functions $\Inputs \to \States$ (see below for some explanation of this notation), with transitions $\transitionCoalg_\moore^\States = \langle \transitionCoalgOut_\moore^\States, \transitionCoalgTr_\moore^\States \rangle$ given by
        \begin{align}
            \transitionCoalgOut_\moore^\States & : \States \to \Observations \nonumber \\
            \transitionCoalgTr_\moore^\States & : \States \to \States^\Inputs,
        \end{align}
        and
        \item as morphisms between two coalgebras $(\States, \transitionCoalg_\moore^\States)$ and $(\StatesPrime, \transitionCoalg_\moore^\StatesPrime)$ with the same inputs and outputs, coalgebra homomorphisms in the form of functions $\homomorphism: \States \to \StatesPrime$ (since inputs and outputs are the same for the two systems, there are simple identity functions between them) that make the following diagram commute
        \begin{equation}
            % https://q.uiver.app/#q=WzAsNCxbMCwwLCJTIl0sWzAsMiwiTyBcXHRpbWVzIFNeSSJdLFsyLDAsIlQiXSxbMiwyLCJPIFxcdGltZXMgVF5JIl0sWzAsMSwiZiIsMl0sWzAsMiwiXFxwaGkiXSxbMiwzLCJnIl0sWzEsMywiXFx0ZXh0e2lkfV9PIFxcdGltZXMgXFxwaGleSSIsMl1d
            \begin{tikzcd}
                \States && \StatesPrime \\
                \\
                {\Observations \times \States^\Inputs} && {\Observations \times \StatesPrime^\Inputs}
                \arrow["\transitionCoalg_\moore^\States"', from=1-1, to=3-1]
                \arrow["\homomorphism", from=1-1, to=1-3]
                \arrow["\transitionCoalg_\moore^\StatesPrime", from=1-3, to=3-3]
                \arrow["{\id_\Observations \times (\homomorphism)^\Inputs}"', from=3-1, to=3-3]
            \end{tikzcd}
        \end{equation}
    \end{itemize}
    
    The notation for the set of all functions $\Inputs \to \States$, $\States^\Inputs = \{f \mid f: \Inputs \to \States\}$, implies that $\transitionCoalgTr_\moore^\States$ sends an element $\states \in \States$ to a function $f = \transitionCoalgTr_\moore^\States(\states): \Inputs \to \States$ assigning a next state $\tilde{\states} \coloneqq \transitionCoalgTr^\States_\moore(\states)(\inputs) \in \States$ to each input $\inputs \in \Inputs$, see~\cite{ruttenMethodCoalgebraExercises2019}.
    Importantly, as noted by~\cite{ruttenUniversalCoalgebraTheory2000} and references therein, there is a bijection $\{h \mid h: \Inputs \times \States \to \States\} \cong \{g \mid g: \States \to \States^\Inputs\}$ (cf. ``currying'')~\footnote{Note that maps $\Inputs \times \States \to \States$ are not of type $\States \to \type(\States)$, while maps of type $\States \to \States^\Inputs$ are, and so are well defined coalgebras.}, and therefore $\{h \mid h: \Inputs \times \States \to \Observations \times \States\} \cong \{h \mid h: \States \to \Observations \times \States^\Inputs\}$.~\footnote{This is true for Moore machines, but not for all kinds of open systems, e.g. Mealy machines where $\Observations$ also depends on $\Inputs$.}
\end{example}

%
% ----------------------------------

\subsubsection{Non-deterministic systems as coalgebras}
Nondeterministic automata provide an example of nondeterministic open discrete-time dynamical systems.
These systems have a non-trivial branching: for each state, there is a set of \emph{possible} next states, and this set can be empty, meaning that from a state there are no transition allowed.
Their transition types can also be of different kinds (closed, open, with or without a terminal state, etc.), but here we focus exclusively on open systems (Moore machines) that include final states (states from which there are no possible transitions). 
This example is also relevant for the next section, where we will introduce yet another type of branching: a probabilistic one.
All these examples can be viewed as special cases of ``structured Moore machines''~\cite{silvaGeneralizingDeterminizationAutomata2013}, where the transition type (with inputs and outputs, specified by a fixed functor, $F$) is the same, while the branching (explained in terms of a monad, $T$) can be different.
To define possibilistic Moore machines, we recall the definition of the finite powerset functor.
\footnote{The finite version of this functor has especially nice properties~\cite{jacobsIntroductionCoalgebraMathematics2017}, and is in practice more often adopted in place of its infinite counterpart.}
\begin{definition}[Finite powerset functor]
    The finite powerset functor, $\PowerFinite : \catset \rightarrow \catset$ is defined as
    \begin{align}
        \PowerFinite(X) = \left\{U \subseteq X \mid U \text{ is finite} \right\}.
    \end{align}
    For a function $\function: X \rightarrow Y$, the (pushforward) map $\PowerFinite(\function): \PowerFinite(X) \rightarrow \PowerFinite(Y)$ is defined as
    \begin{align}
        \PowerFinite(\function)(\actions) & = \function[U],
    \end{align}
    where $\function[U] = \{ \function(\actions) \mid u \in U\}$.
\end{definition}

Using this functor, we can give the following definition.

\begin{example}[Category of nondeterministic Moore machines as coalgebras]
    \label{ex:nondetMoore}
    The category of nondeterministic Moore machines as coalgebras $\Coalg(\powermoore)$, with coalgebra type given by the functor $\powermoore : \catset \to \catset$ such that $\powermoore(\States) = \PowerFinite(\Observations) \times \PowerFinite(\States)^\Inputs$, has 
    \begin{itemize}
        \item as objects, coalgebras of the form $(\States, \transitionCoalg_\powermoore^\States : \States \to \PowerFinite(\Observations) \times \PowerFinite(\States)^\Inputs)$, with transitions $\transitionCoalg_\powermoore^\States = \langle \transitionCoalgTr_\powermoore^\States, \transitionCoalgOut_\powermoore^\States \rangle$ given by
        \begin{align}
            \transitionCoalgOut_\powermoore^\States & : \States \to \PowerFinite(\Observations) \nonumber \\
            \transitionCoalgTr_\powermoore^\States & : \States \to \PowerFinite(\States)^\Inputs,
        \end{align}
        and
        \item as morphisms between two coalgebras $(\States, \transitionCoalg_\powermoore^\States)$ and $(\StatesPrime, \transitionCoalg_\powermoore^\StatesPrime)$ with the same inputs and outputs, coalgebra homomorphisms in the form of functions $\homomorphism: \States \to \StatesPrime$ (since inputs and outputs are the same for the two systems, there are simple identity functions between them) that make the following diagram commute
        \begin{equation}
            % https://q.uiver.app/#q=WzAsNCxbMCwwLCJTIl0sWzAsMiwiTyBcXHRpbWVzIFNeSSJdLFsyLDAsIlQiXSxbMiwyLCJPIFxcdGltZXMgVF5JIl0sWzAsMSwiZiIsMl0sWzAsMiwiXFxwaGkiXSxbMiwzLCJnIl0sWzEsMywiXFx0ZXh0e2lkfV9PIFxcdGltZXMgXFxwaGleSSIsMl1d
            \begin{tikzcd}
                \States &&& \StatesPrime \\
                \\
                {\PowerFinite(\Observations) \times \PowerFinite(\States)^\Inputs} &&& {\PowerFinite(\Observations) \times \PowerFinite(\StatesPrime)^\Inputs}
                \arrow["\transitionCoalg_\powermoore^\States"', from=1-1, to=3-1]
                \arrow["\homomorphism", from=1-1, to=1-4]
                \arrow["\transitionCoalg_\powermoore^\States", from=1-4, to=3-4]
                \arrow["{\PowerFinite(\id_\Observations) \times \PowerFinite(\homomorphism)^\Inputs}"', from=3-1, to=3-4]
            \end{tikzcd}
        \end{equation}
    \end{itemize}
\end{example}

We note that this formulation is closely related to the idea of \emph{possibilistic}~\cite{aubinViabilityTheoryNew2011} systems, defined for the \emph{nonempty} powerset monad given by $\PowerFinite^+(X) = \PowerFinite(X) \setminus \{\varnothing\}$, which exclude the case where there can be no possible transition from a given state.

%
%
% -----------------------------------------------------------------------------------------
% -----------------------------------------------------------------------------------------

\subsection{Comparing processes}
\label{sec:equivalence}
A core feature of the coalgebraic approach to dynamical systems is its focus on maps between systems, placing them at the center of the theory’s development.
These maps are relevant, for instance, in the analysis of concurrent processes in theoretical computer science~\cite{ruttenUniversalCoalgebraTheory2000, sangiorgiAdvancedTopicsBisimulation2011}, where a primary goal is to define notions of equivalence for processes based on their observable behaviour, contrasting with standard concepts of equivalence based on structural similarity of different processes~\cite{sangiorgiIntroductionBisimulationCoinduction2011}.
They are also of interest in other fields, where they appear in specialised forms such as ``homomorphisms''~\cite{ravindranSMDPHomomorphismsAlgebraic2003, rosasAIVatFundamental2025}, ``coarse-grainings''~\cite{noidMultiscaleCoarsegrainingMethod2008}, ``variable aggregation''~\cite{simonAggregationVariablesDynamic1961}, ``state aggregation''~\cite{renStateAggregationMarkov2003}, ``lumpability''~\cite{kemenyFiniteMarkovChains1969}, ``model reduction''~\cite{moorePrincipalComponentAnalysis1981} or ``dynamical consistency''~\cite{cainesHierarchicalLatticesFinite1995}, among others.

These specialised maps can be traced back to the idea of a \emph{model}, in particular, to what it means for a system to model another system.
The standard definition of model relies on epimorphisms, a generalisation of surjectivity for functions between sets. 
In the case of coalgebras, it tells when a coalgebra can be specified as a coarse-graining of another.

\begin{definition}[Models in coalgebras]
    \label{def:model}
    Let $\Coalg(\type)$ be a category of coalgebras.
    A homomorphism of coalgebras $\homomorphism : \States \to \StatesPrime$ from $(\States, \transitionCoalg^\States : \States \to \type(\States))$ to $(\StatesPrime, \transitionCoalg^\StatesPrime : \StatesPrime \to \type(\StatesPrime))$ is an epimorphism in the category $\Coalg(\type)$ if and only if $\homomorphism$ is a surjective function between the underlying sets
   ~\cite[Theorem 3.3.4]{jacobsIntroductionCoalgebraMathematics2017}.
   Following the convention adopted in~\cite{baltieriBayesianInterpretationInternal2025}, an epimorphism of coalgebras $\homomorphism$ is called a \emph{model}, whereas $(\States, \transitionCoalg^\States)$ is called the \emph{referent}, i.e. what the model refers to, and $(\StatesPrime, \transitionCoalg^\StatesPrime)$ the \emph{referrer}, i.e. what refers to the model.
\end{definition}

Intuitively, this definition states that different elements of the first coalgebra, $(\States, \transitionCoalg^\States)$, are mapped to the same element of the second coalgebra, $(\StatesPrime, \transitionCoalg^\StatesPrime)$.
More precisely, the existence of a model $\homomorphism: \States \to \StatesPrime$ implies that for each state $\statesPrime \in \StatesPrime$, there exists a set of states $\invhomomorphism(\statesPrime) \in \States$ of the referent $(\States, \transitionCoalg^\States)$, called the \emph{fibre} of $\statesPrime$, which represents a subset of elements of $\States$ that are \emph{indistinguishable} from the perspective of the simpler coalgebra, the referrer $(\StatesPrime, \transitionCoalg^\StatesPrime)$, as they all map to the same element $\statesPrime \in \StatesPrime$ via the surjective function $\homomorphism$.
Furthermore, as $\statesPrime \in \StatesPrime$ varies along the transition function $\transitionCoalgPrime$, this variation is consistent with the variation described by the function $\transitionCoalg$ for each element $\states$ of the fibre $\invhomomorphism(\statesPrime)$ of $\statesPrime$.

An equivalent definition, see~\cite[Theorem 3.3.4]{jacobsIntroductionCoalgebraMathematics2017}, can be given via the use of \emph{bisimulation equivalences}, building on the concepts of spans and relations.

\begin{definition}[Spans and relations between sets]
    \label{def:span}
    A span between the sets $X$ and $Y$ is a triple $(V, p_1, p_2)$ where $V$ is a set and $p_1 : V \to X$ and $p_2 : V \to Y$ are two functions with the same domain, $V$.
    The pair $(x, y) \in X \times Y$ is related by $(V, p_1, p_2)$ if there exists a $v \in V$ such that $p_1(v) = x$ and $p_2(v) = y$.
    A relation $R$ is a \emph{jointly monic span}, i.e. a span where $p_1, p_2$ are jointly a monomorphism (an injective function in $\catset$), $R \xrightarrow{(p_1, p_2)} X \times Y$, given by $v \mapsto (p_1(v), p_2(v))$, or in other words, given any two functions $f, g : W \to V$, $f \comp p_1 = g \comp p_1$ and $f \comp p_2 = g \comp p_2$ imply that $f = g$.
\end{definition}

\begin{definition}[Bisimulation equivalence]
    \label{def:bisimulationEquivalence}
    Given an {$\type$-coalgebra} $(\States, \transitionCoalg^\States)$ an equivalence relation $\Bisim \subseteq \States \times \States$ is said to be an $\type$-\emph{bisimulation equivalence}~\cite{ruttenUniversalCoalgebraTheory2000} if there exists an $\type$-coalgebra structure $\transitionBisim^\Bisim: \Bisim \to \type(\Bisim)$ such that the following diagram commutes
    \begin{equation}
        % https://q.uiver.app/#q=WzAsNixbMiwwLCJSIl0sWzIsMiwiRihSKSJdLFs0LDAsIlQiXSxbNCwyLCJGKFQpIl0sWzAsMCwiUyJdLFswLDIsIkYoUykiXSxbMCwxLCJcXGdhbW1hIiwwLHsic3R5bGUiOnsiYm9keSI6eyJuYW1lIjoiZGFzaGVkIn19fV0sWzAsMiwiXFxwaV8yIl0sWzIsMywiZyJdLFsxLDMsIkYgKFxccGlfMikiLDJdLFswLDQsIlxccGlfMSIsMl0sWzEsNSwiRiAoXFxwaV8xKSJdLFs0LDUsImYiLDJdLFswLDEsIlxcZXhpc3RzIiwyLHsic3R5bGUiOnsiYm9keSI6eyJuYW1lIjoibm9uZSJ9LCJoZWFkIjp7Im5hbWUiOiJub25lIn19fV1d
        \begin{tikzcd}
            \States && \Bisim && \States \\
            \\
            {\Func(\States)} && {\Func(\Bisim)} && {\Func(\States)}
            \arrow["\transitionBisim^\Bisim", dashed, from=1-3, to=3-3]
            \arrow["{\pi_\States}", from=1-3, to=1-5]
            \arrow["\transitionCoalg^\States", from=1-5, to=3-5]
            \arrow["{\Func(\pi_\States)}"', from=3-3, to=3-5]
            \arrow["{\pi_{\States}}"', from=1-3, to=1-1]
            \arrow["{\Func(\pi_{\States})}", from=3-3, to=3-1]
            \arrow["\transitionCoalg^\States"', from=1-1, to=3-1]
            \arrow["\exists"', draw=none, from=1-3, to=3-3]
        \end{tikzcd}
    \end{equation}

    i.e. such that the projection $\pi_{\States}$ is an coalgebra homomorphism.
    More generally, a bisimulation equivalence is at times defined as a span of coalgebras, i.e. a span $(\Bisim, \pi_{\States}, \pi_{\States})$ between the underlying sets, $\States$ and $\States$ (itself), that makes the above diagram commute~\cite{jacobsIntroductionCoalgebraMathematics2017}.
    % In this context, a bisimulation equivalence is a special kind of relation between coalgebras in the sense that it invokes a ``jointly monic span'', i.e. a relation instead of a span.
\end{definition}

To get an intuition for how this relates to the modelling perspective of~\cref{def:model}, recall that a model is an epimorphism of coalgebras, i.e. a surjective function between the underlying sets of two coalgebras $(\States, \transitionCoalg^\States)$ and $(\StatesPrime, \transitionCoalg^\StatesPrime)$.
Recall also that every surjective function $f : A \to B$ induces an equivalence relation on $A$, $R \subseteq A \times A$ and conversely, every equivalence relation $R$ on $A$ induces a quotient mapping $f_R : A \to A/R$, which is surjective, where $A/R$ is the quotient set with elements equivalence classes of elements of $A$.
In this sense, an equivalence relation is to a surjective function what a bisimulation equivalence is to a coalgebra epimorphism: a bisimulation equivalence is an equivalence relation $\Bisim$ on $\States$ with extra structure, i.e. an equivalence relation that preserves coalgebra transitions, and a coalgebra epimorphism is a surjective function between the underlying sets of two coalgebras, $\States$ and $\StatesPrime$, with extra structure, i.e. a surjective function that preserves coalgebra transitions.
This interpretation has recently received increasing attention in machine and reinforcement learning, see for instance~\cite{zhangLearningCausalState2021, zhangStateAbstractionsGeneralization2021, rosasAIVatFundamental2025}, where bisimulation equivalences are simply referred to as bisimulations.

Following~\cite{devinkBisimulationProbabilisticTransition1999, desharnaisBisimulationLabelledMarkov2002, jacobsIntroductionCoalgebraMathematics2017}, we extend the above definition to relations (not equivalence relations) of the form $\Bisim \subseteq \States \times \StatesPrime$ between different sets of states $\States$ and $\StatesPrime$ and build a definition of bisimulation between \emph{different} processes formalised as coalgebras. 

\begin{definition}[Bisimulation]
    \label{def:bisimulation}
    Given two {$\type$-coalgebras} $(\States, \transitionCoalg^\States), (\StatesPrime, \transitionCoalg^\StatesPrime)$, a relation $\Bisim$ is said to be an $\type$-\emph{bisimulation}~\cite{ruttenUniversalCoalgebraTheory2000} between $(\States, \transitionCoalg^\States)$ and $(\StatesPrime, \transitionCoalg^\StatesPrime)$ if there exists an $\type$-coalgebra structure $\transitionBisim^\Bisim: \Bisim \to \type(\Bisim)$ such that the following diagram commutes
    \begin{equation}
        % https://q.uiver.app/#q=WzAsNixbMiwwLCJSIl0sWzIsMiwiRihSKSJdLFs0LDAsIlQiXSxbNCwyLCJGKFQpIl0sWzAsMCwiUyJdLFswLDIsIkYoUykiXSxbMCwxLCJcXGdhbW1hIiwwLHsic3R5bGUiOnsiYm9keSI6eyJuYW1lIjoiZGFzaGVkIn19fV0sWzAsMiwiXFxwaV8yIl0sWzIsMywiZyJdLFsxLDMsIkYgKFxccGlfMikiLDJdLFswLDQsIlxccGlfMSIsMl0sWzEsNSwiRiAoXFxwaV8xKSJdLFs0LDUsImYiLDJdLFswLDEsIlxcZXhpc3RzIiwyLHsic3R5bGUiOnsiYm9keSI6eyJuYW1lIjoibm9uZSJ9LCJoZWFkIjp7Im5hbWUiOiJub25lIn19fV1d
        \begin{tikzcd}
            \States && \Bisim && \StatesPrime \\
            \\
            {\Func(\States)} && {\Func(\Bisim)} && {\Func(\StatesPrime)}
            \arrow["\transitionBisim^\Bisim", dashed, from=1-3, to=3-3]
            \arrow["{\pi_\StatesPrime}", from=1-3, to=1-5]
            \arrow["\transitionCoalg^\StatesPrime", from=1-5, to=3-5]
            \arrow["{\Func(\pi_\StatesPrime)}"', from=3-3, to=3-5]
            \arrow["{\pi_\States}"', from=1-3, to=1-1]
            \arrow["{\Func(\pi_\States)}", from=3-3, to=3-1]
            \arrow["\transitionCoalg^\States"', from=1-1, to=3-1]
            \arrow["\exists"', draw=none, from=1-3, to=3-3]
        \end{tikzcd}
    \end{equation}
    i.e. such that $\pi_\States, \pi_\StatesPrime$ are coalgebra homomorphisms.
    Again, we can generalise this definition to state that a bisimulation is a span of coalgebras.
\end{definition}

Unlike the case of bisimulation equivalences, i.e. equivalence relations of type $R \subseteq A \times A$ that correspond to models and coalgebra epimorphism given in~\cref{def:model}, for relations of type $R \subseteq A \times B$ the correspondence is less obvious. This is primarily because relations of type $R \subseteq A \times B$ do not simply induce a surjective function, see also Discussion and~\cite{moroniClassificationBisimilaritiesGeneral2024}.

An alternative approach to understanding the more general notion of bisimulation is through the lenses of \emph{behavioural equivalence}.
However, to establish a rigorous notion of behavioural equivalence, we first need to define cocongruences, kernel bisimulations and final coalgebras. Although in some other works, ``behavioural equivalence'', ``cocongruence'' and ``kernel bisimulation'' are used interchangeably, here we adopt the formal characterisation given in~\cite{statonRelatingCoalgebraicNotions2011}, which distinguishes these notions by, roughly, stating that behavioural equivalence is a kernel bisimulation that uses a final coalgebra, and kernel bisimulation is a cocongruence with an associated (pullback) relation. See also~\cite[Chapter 3.5]{bacciGeneralizedLabelledMarkov2013} for a detailed breakdown.

The definition of \emph{cocongruence} turns out to be equivalent to that of bisimulation in our setup, i.e. using only weak pullback-preserving functors on $\catset$ as the base category~\cite{barbosaCoalgebraWorkingSoftware2022} and~\cite[Theorem 4.5.3]{jacobsIntroductionCoalgebraMathematics2017}.
Cocongruences in~\cite{kurzLogicsCoalgebrasApplications2001} (or behavioural equivalence in~\cite{jacobsIntroductionCoalgebraMathematics2017}) build on the dual\footnote{Formal duality, in the sense of category theory, i.e. \emph{arrow reversal}.} of a relation (a so-called ``corelation'') and more generally, the dual of a span (a ``cospan'').
\begin{definition}[Cospans and corelations between sets]
    \label{def:cospan}
    A cospan between the sets $X$ and $Y$ is a triple $(U, i_1, i_2)$ where $U$ is a set and $i_1 : X \to U$ and $i_2 : Y \to U$ are two functions with the same codomain, $U$.
    The pair $(x, y) \in X \times Y$ is identified by $(U, i_1, i_2)$ if $i_1(x) = i_2(y)$.
    A corelation $C$ is a \emph{jointly epic cospan}, i.e. a cospan where $i_1, i_2$ are jointly an epimorphism (a surjective function in $\catset$), $X + Y \xrightarrow{(i_1, i_2)} C$, given by $x \mapsto i_1(x)$ and $y \mapsto i_2(y)$, or in other words given any two functions $h, k : U \to T$, $i_1 \comp h = i_1 \comp k$ and $i_2 \comp h = i_2 \comp k$ imply that $h = k$.
\end{definition}

\begin{definition}[Cocongruence]
    \label{def:cocongruence}
    Given two {$\type$-coalgebras} $(\States, \transitionCoalg^\States), (\StatesPrime, \transitionCoalg^\StatesPrime)$, a corelation $C$ is said to be a cocongruence~\cite{kurzLogicsCoalgebrasApplications2001, bacciGeneralizedLabelledMarkov2013} between $(\States, \transitionCoalg^\States)$ and $(\StatesPrime, \transitionCoalg^\StatesPrime)$ if there exists an $\type$-coalgebra structure $\transitionBisim^C: C \to \type(C)$ such that the following diagram commutes:
    \begin{equation}
        % https://q.uiver.app/#q=WzAsNixbMiwwLCJSIl0sWzIsMiwiRihSKSJdLFs0LDAsIlQiXSxbNCwyLCJGKFQpIl0sWzAsMCwiUyJdLFswLDIsIkYoUykiXSxbMCwxLCJcXGdhbW1hIiwwLHsic3R5bGUiOnsiYm9keSI6eyJuYW1lIjoiZGFzaGVkIn19fV0sWzIsMywiZyJdLFs0LDUsImYiLDJdLFswLDEsIlxcZXhpc3RzIiwyLHsic3R5bGUiOnsiYm9keSI6eyJuYW1lIjoibm9uZSJ9LCJoZWFkIjp7Im5hbWUiOiJub25lIn19fV0sWzQsMCwicl8xIl0sWzIsMCwicl8yIiwyXSxbNSwxLCJGKHJfMSkiLDJdLFszLDEsIkYocl8yKSJdXQ==
        \begin{tikzcd}
        	\States && C && \StatesPrime \\
        	\\
        	{\type(\States)} && {\type(C)} && {\type(\StatesPrime)}
        	\arrow["{r_\States}", from=1-1, to=1-3]
        	\arrow["\transitionCoalg^\States"', from=1-1, to=3-1]
        	\arrow["\transitionBisim^C", dashed, from=1-3, to=3-3]
        	\arrow["\exists"', draw=none, from=1-3, to=3-3]
        	\arrow["{r_\StatesPrime}"', from=1-5, to=1-3]
        	\arrow["\transitionCoalg^\StatesPrime", from=1-5, to=3-5]
        	\arrow["{\type(r_\States)}"', from=3-1, to=3-3]
        	\arrow["{\type(r_\StatesPrime)}", from=3-5, to=3-3]
        \end{tikzcd}
    \end{equation}
    i.e. such that $r_\States, r_\StatesPrime$ are coalgebra homomorphisms. More generally, cocongruence can be characterised as a cospan of coalgebras, with the above as a special case.
\end{definition}

A \emph{kernel bisimulation} is, in this context, a relation $R$ associated, when it exists, to a cocongruence.
\begin{definition}[Kernel bisimulation]
    \label{def:kernelBisimulation}
    Given a cocongruence between two {$\type$-coalgebras} $(\States, \transitionCoalg^\States), (\StatesPrime, \transitionCoalg^\StatesPrime)$, a relation $R \subseteq \States \times \StatesPrime$ is a kernel bisimulation if it is a pullback of the cospan $\States \to C \gets \StatesPrime$, i.e. if there exist morphisms $s_\States: R \to \States$ and $s_\StatesPrime : R \to \StatesPrime$ such that the following diagram commutes:
    \begin{equation}
        \begin{tikzcd}
                && R && \\
        	\States && C && \StatesPrime \\
        	\\
        	{\type(\States)} && {\type(C)} && {\type(\StatesPrime)}
                \arrow["{s_\States}"', from=1-3, to=2-1]
                \arrow["{s_\StatesPrime}", from=1-3, to=2-5]
        	\arrow["{r_\States}", from=2-1, to=2-3]
        	\arrow["\transitionCoalg^\States"', from=2-1, to=4-1]
        	\arrow["\transitionBisim^C", from=2-3, to=4-3]
        	\arrow["\exists"', draw=none, from=2-3, to=4-3]
        	\arrow["{r_\StatesPrime}"', from=2-5, to=2-3]
        	\arrow["\transitionCoalg^\StatesPrime", from=2-5, to=4-5]
        	\arrow["{\type(r_\States)}"', from=4-1, to=4-3]
        	\arrow["{\type(r_\StatesPrime)}", from=4-5, to=4-3]
        \end{tikzcd}
    \end{equation}
    In a category of coalgebras for a functor preserving weak pullbacks, as is the case for all the functors considered in this work, with a base category with pullbacks (such as $\catset$), cocongruence and kernel bisimulations imply each other, i.e. every cocongruence has an associated kernel bisimulation.
\end{definition}

A \emph{final coalgebra}, when it exists, is the final or terminal object in a category $\Coalg(\type)$.
\begin{definition}[\textbf{Final coalgebra}]
    \label{def:finalCoalgebra}
    An $\type$-coalgebra $(\StatesFinal, \transitionCoalgFinal^\StatesFinal)$ is final in the category $\Coalg(\type)$ if for any $\type$-coalgebra $(\States, \transitionCoalg^\States)$ there exists a unique $\type$-homomorphism $\beh_\States : (\States, \transitionCoalg^\States) \to (\StatesFinal, \transitionCoalgFinal)$.
    Graphically, this is equivalent to the following diagram commuting:
    \begin{equation}
        % https://q.uiver.app/#q=WzAsNCxbMCwwLCJBIl0sWzAsMiwiRkEiXSxbMiwyLCJGQiJdLFsyLDAsIkIiXSxbMCwxLCJmIiwyXSxbMSwyLCJGYSIsMix7InN0eWxlIjp7ImJvZHkiOnsibmFtZSI6ImRhc2hlZCJ9fX1dLFswLDMsImEiLDAseyJzdHlsZSI6eyJib2R5Ijp7Im5hbWUiOiJkYXNoZWQifX19XSxbMywyLCJnIl1d
        \begin{tikzcd}
        	\States && \StatesFinal \\
        	\\
        	\Func(\States) && \Func(\StatesFinal)
        	\arrow["\beh_\States", dashed, from=1-1, to=1-3]
        	\arrow["\transitionCoalg^\States"', from=1-1, to=3-1]
        	\arrow["\transitionCoalgFinal^\StatesFinal", from=1-3, to=3-3]
        	\arrow["\Func(\beh_\States)"', dashed, from=3-1, to=3-3]
        \end{tikzcd}
    \end{equation}
\end{definition}

A final coalgebra is said to capture the behaviour of a coalgebra~\cite{ruttenUniversalCoalgebraTheory2000, jacobsIntroductionCoalgebraMathematics2017}.
More precisely, the elements of a final coalgebra (when it exists) are the possible observable behaviours of all objects (including itself) of a given category of coalgebras.
We recall from~\cref{sec:synchronisation} that, in the theory of coalgebras, behaviour was informally defined as ``outputs of a system over time''.
In coalgebraic terms, however, the rigorous definition of behaviour is more complicated, and depends strictly on the type of functor used to build coalgebras: behaviours could correspond to traces (repeated applications of the coalgebra transition map), trees, distributions, etc.
For the purposes of this paper, we will only consider a couple of standard examples relevant to predictive processing in~\cref{sec:PPcoalgebras}, and refer the reader to standard treatments such as~\cite{ruttenUniversalCoalgebraTheory2000, jacobsIntroductionCoalgebraMathematics2017} for a more in depth discussion of final coalgebras and their semantics. 

\begin{example}
    For the category of closed systems in~\cref{def:ClosedSystems}, the final coalgebra is trivial, it is the one element set $\{\cdot\}$ since all systems look the same from an external perspective, i.e. nothing can be observed because these systems have no outputs.
\end{example}

Next, we look at the case of deterministic Moore machines.

\begin{example}
    Let $\Coalg(\moore)$ be the category of Moore machines as coalgebras from~\cref{ex:MooreMachines}.
    The final coalgebra in $\Coalg(\moore)$ is given by
    \begin{align}
        (\Observations^{\Inputs^\ast}, \transitionCoalgFinal^{\Observations^{\Inputs^\ast}}: \Observations^{\Inputs^\ast} \to \Observations \times (\Observations^{\Inputs^\ast})^\Inputs,
    \end{align}
    where the notation $^\ast$ is used to represent lists~\cite{ruttenUniversalCoalgebraTheory2000, jacobsIntroductionCoalgebraMathematics2017}.
    Here, the carrier of the final coalgebra is $\Observations^{\Inputs^\ast}$, and its elements $\in \Observations^{\Inputs^\ast}$ can be understood by defining state transitions from any initial state $\states_0 \in \States$ for a given list of actions of \emph{arbitrary} length $n$, $\langle \inputs_1, \dots, \inputs_n \rangle \in \Inputs^\ast$.
    Using these, we can take $n$ steps from the initial state $\states_0$, i.e. $\transitionCoalgTr_\moore(\dots \transitionCoalgTr_\moore(\states_0)(\inputs_1)) \dots)(\inputs_n)$, and obtain an observation for each list $\transitionCoalgOut_\moore(\dots \transitionCoalgTr_\moore(\states_0)(\inputs_1)) \dots)(\inputs_n)$~\cite[Sec. 2.2.3]{jacobsIntroductionCoalgebraMathematics2017}, thus obtaining trees rooted in some initial state $\states_0$ with inputs as edges between nodes represented by the possible outputs given those edges.
\end{example}

We now combine the definitions of kernel bisimulation, which can be built from all cocongruences in categories of coalgebras using only weak pullback-preserving functors on $\catset$ as the base category, and final coalgebra to obtain the following~\cite{statonRelatingCoalgebraicNotions2011}.
\begin{definition}[\textbf{Behavioural equivalence}]
    \label{def:behaviouralEquivalence}
    Given two {$\type$-coalgebras} $(\States, \transitionCoalg^\States), (\StatesPrime, \transitionCoalg^\StatesPrime)$, behavioural equivalence between them is a kernel bisimulation $R$ (\cref{def:kernelBisimulation}) for a cocongruence $(\StatesFinal, \transitionBisim^\StatesFinal: \StatesFinal \to \type(\StatesFinal))$ (\cref{def:cocongruence}) where $(\StatesFinal, \transitionBisim^\StatesFinal)$ is the final coalgebra (\cref{def:finalCoalgebra}).

    Two states $\states \in \States, \statesPrime \in \StatesPrime$ are \emph{behaviourally equivalent}, i.e. $(\states, \statesPrime) \in R$, if for $r_\States : \States \to C, r_\States: \StatesPrime \to C$ (see~\Cref{def:cocongruence,def:kernelBisimulation}), $r_\States(\states) = r_\StatesPrime(\statesPrime)$ (see~\cite[Theorem 3.3.3]{jacobsIntroductionCoalgebraMathematics2017} for polynomial functors and its generalisation to include the finite powerset functor and the distribution functor~\cite[Theorem 4.5.3]{jacobsIntroductionCoalgebraMathematics2017}).
\end{definition}

In the next section, we will use behavioural equivalence to provide a structural description of the core parts of predictive processing, and free energy minimisation, under a coalgebraic framework.
As we shall see, given the level of abstraction we reached, we will only need to apply a few minor changes to definitions introduced above to understand the relation that ought to be in place between generative process and generative model for a successful predictive processing agent.
Although we do not focus on any specific algorithmic implementation, our final discussion will provide a connection to existing work on both exact and approximate bisimulations, showing how some of the ideas introduced in the next section could be implemented in future work.

%
%
%
% -----------------------------------------------------------------------------------------
% -----------------------------------------------------------------------------------------
% -----------------------------------------------------------------------------------------

\section{Predictive processing in coalgebraic terms}
\label{sec:PPcoalgebras}
\subsection{Generative model and generative process as coalgebras}
In~\cref{sec:activeInferenceStructure}, we saw that, in the predictive processing literature, the terms generative process and generative model have been used as labels for probabilistic processes that represent the ground truth of the environment, and a model of it whose updates are encoded in the agent's brain states, respectively~\cite{dacostaActiveInferenceDiscrete2020, mannellaActiveInferenceWhiskers2021, parrActiveInferenceFree2022, sajidActiveInferenceDemystified2021}. 
These probabilistic processes are usually presented as POMDPs (see~\cref{def:POMDP}), and in the literature on coalgebras, correspond to a probabilistic version of Moore machines (cf.~\cref{ex:MooreMachines}).
As in the case of~\cref{ex:nondetMoore}, these constitute another example of branching (probabilistic) given the same transition type (with inputs and outputs) of structured Moore machines~\cite{silvaGeneralizingDeterminizationAutomata2013}.
To define them, we first recall the following.
\begin{definition}[\textbf{Distribution functor}]
    \label{def:distFunctor}
    The distribution functor for discrete probability, $\Dist : \catset \rightarrow \catset$ is defined as:
    \begin{align}
        \Dist(X) = \left\{\dist: X \rightarrow [0,1] \,\middle\vert\, \supp(\dist) \text { is finite and } \sum_x \dist(x) = 1 \right\},
    \end{align}
    where $[0,1] \subseteq \Reals$ is the unit interval of real numbers, and $\supp(\dist) \subseteq X$ is the support of the distribution, i.e. the finite subset of $x \in X$ where $\dist(x) \neq 0$.
    For a function $\function: X \rightarrow Y$ the map $\Dist(\function): \Dist(X) \rightarrow \Dist(Y)$ (the pushforward of $\dist$ along $\function$) is defined, for any distribution $\dist \in \Dist(X)$ and any element $y \in Y$, as:
    \begin{align}
        \Dist(\function)(\dist) (y) & = \sum_{x \in \function^{-1}(y)} \dist(x) = \sum_x \{\dist(x) \mid x \in \supp(\dist) \text { with } \function(x)=y\} .
    \end{align}
    \label{def:distributionFunctor}
\end{definition}

Using this, we can define probabilistic Moore machines and interpret them as POMDPs in coalgebraic terms. Note that we adopt a common simplifying assumption stating that observations and transitions to the next state are independent. For more general treatments not involving coalgebras, see for instance~\cite{biehlInterpretingSystemsSolving2022, virgoUnifilarMachinesAdjoint2023, rosasAIVatFundamental2025}.
\begin{definition}[\textbf{Category of POMDPs as coalgebras (probabilistic Moore machines in~\cite{silvaGeneralizingPowersetConstruction2010})}]
    \label{def:categoryPOMDP}

    The category of POMDPs as coalgebras $\Coalg(\pomdp)$, with coalgebra type given by the functor $\pomdp : \catset \to \catset$ such that $\pomdp(\States) = \Dist(\Observations) \times \Dist(\States)^\Actions$, has 
    \begin{itemize}
        \item as objects, coalgebras of the form $(\States, \transitionCoalg_\pomdp^\States : \States \to \Dist(\Observations) \times \Dist(\States)^\Actions)$, with transitions $\transitionCoalg_\pomdp^\States = \langle \transitionCoalgOut_\pomdp^\States, \transitionCoalgTr_\pomdp^\States \rangle$ given by
        \begin{align}
            \transitionCoalgOut_\pomdp^\States & : \States \to \Dist(\Observations) \nonumber \\
            \transitionCoalgTr_\pomdp^\States & : \States \to \Dist(\States)^\Actions,
        \end{align}
        and
        \item as morphisms between two coalgebras $(\States, \transitionCoalg_\pomdp^\States)$ and $(\StatesPrime, \transitionCoalg_\pomdp^\States
        )$ with the same inputs/actions and outputs/observations, coalgebra homomorphisms in the form of functions $\homomorphism: \States \to \StatesPrime$ (since inputs and outputs are the same for the two systems, there are simple identity functions between them, indicated by $\id_\Observations$ for observations, and by having the same $\Actions$ on both coalgebras for actions) that make the following diagram commute:
        \begin{equation}
            % https://q.uiver.app/#q=WzAsNCxbMCwwLCJTIl0sWzAsMiwiTyBcXHRpbWVzIFNeSSJdLFsyLDAsIlQiXSxbMiwyLCJPIFxcdGltZXMgVF5JIl0sWzAsMSwiZiIsMl0sWzAsMiwiXFxwaGkiXSxbMiwzLCJnIl0sWzEsMywiXFx0ZXh0e2lkfV9PIFxcdGltZXMgXFxwaGleSSIsMl1d
            \begin{tikzcd}
                \States &&& \StatesPrime \\
                \\
                {\Dist(\Observations) \times \Dist(\States)^\Actions} &&& {\Dist(\Observations) \times \Dist(\StatesPrime)^\Actions}
                \arrow["\transitionCoalg_\pomdp^\States"', from=1-1, to=3-1]
                \arrow["\homomorphism", from=1-1, to=1-4]
                \arrow["\transitionCoalg_\pomdp^\StatesPrime", from=1-4, to=3-4]
                \arrow["{\Dist(\id_\Observations) \times \Dist(\homomorphism)^\Actions}"', from=3-1, to=3-4]
            \end{tikzcd}
        \end{equation}
    \end{itemize}
\end{definition}

Similarly, we define the category of MDPs below, following in this case previous work~\cite{feysLongTermValuesMarkov2018}, but without including explicit rewards.

\begin{definition}[\textbf{Category of MDPs as coalgebras}]
    \label{def:categoryMDP}
    The category of MDPs as coalgebras $\Coalg(\mdp)$, with coalgebra type given by the functor $\mdp : \catset \to \catset$ such that $\mdp(\States) = \Dist(\States)^\Actions$, has 
    \begin{itemize}
        \item as objects, coalgebras of the form $(\States, \transitionCoalg_\mdp^\States : \States \to \Dist(\States)^\Actions)$\footnote{Notice how the relation between partially and fully observable MDPs appears: $\transitionCoalg_\mdp^\States$ is of the same type as $\transitionCoalgTr_\pomdp^\States$.}, and
        \item as morphisms between two coalgebras $(\States, \transitionCoalg_\mdp^\States)$ and $(\StatesPrime, \transitionCoalg_\mdp^\StatesPrime)$ with the same inputs, coalgebra homomorphisms in the form of functions $\homomorphism: \States \to \StatesPrime$ (since inputs are the same, there is once again a simple identity between them) that make the following diagram commute:
        \begin{equation}
            % https://q.uiver.app/#q=WzAsNCxbMCwwLCJTIl0sWzAsMiwiTyBcXHRpbWVzIFNeSSJdLFsyLDAsIlQiXSxbMiwyLCJPIFxcdGltZXMgVF5JIl0sWzAsMSwiZiIsMl0sWzAsMiwiXFxwaGkiXSxbMiwzLCJnIl0sWzEsMywiXFx0ZXh0e2lkfV9PIFxcdGltZXMgXFxwaGleSSIsMl1d
            \begin{tikzcd}
                \States &&& \StatesPrime \\
                \\
                {\Dist(\States)^\Actions} &&& {\Dist(\StatesPrime)^\Actions}
                \arrow["\transitionCoalg_\mdp^\States"', from=1-1, to=3-1]
                \arrow["\homomorphism", from=1-1, to=1-4]
                \arrow["\transitionCoalg_\mdp^\StatesPrime", from=1-4, to=3-4]
                \arrow["{\Dist(\homomorphism)^\Actions}"', from=3-1, to=3-4]
            \end{tikzcd}
        \end{equation}
    \end{itemize}
\end{definition}

\subsection{Comparing generative process and generative model: predictive processing as behavioural equivalence}
\label{sec:behaviouralEquivalencesPP}
In~\cref{sec:synchronisation} we saw how the literature on active inference and predictive processing contains several claims that the process of minimising variational free energy, used to perform approximate Bayesian inference on the environment's states that generate sensory inputs, can be understood in terms of a ``synchronisation'' between generative model and generative process.
This means that, for a particular task, the dynamics of a generative model, implicitly encoded by brain states representing (approximate) Bayesian updates given observations over time, becomes a model, ideally a perfect one, of the generative process.
Here, we provide three candidate, formal definitions of this idea corresponding to three particular forms of behavioural equivalence between generative model and generative process.
We will discuss their implications and possible shortcomings, focusing in the end on what we believe to be the best candidate to reflect a relation between generative model and generative process that goes beyond mere structural similarity, consistent with predictive processing.

%
% -----------------------------------------------------------------------------------------

\subsubsection{Comparing POMDPs}
To start off, we apply directly the definition of behavioural equivalence given in~\cref{def:behaviouralEquivalence} to generative process and generative model in the category of POMDPs (\cref{def:categoryPOMDP}), which, as we know, has a final coalgebra, see~\cite[Section 7]{mossFinalCoalgebrasFunctors2006} and~\cite[Theorem 4.6.9]{jacobsIntroductionCoalgebraMathematics2017}.
As we will see shortly, this has quite strong and perhaps undesirable implications, which are nevertheless important to discuss.
In what follows, we will make extensive use of~\cref{def:behaviouralEquivalence}, but without visualising the relation $R$ (the kernel bisimulation) in our diagrams, since its existence is always implied by our setup, see~\cref{def:kernelBisimulation}.

\begin{definition}[Behavioural equivalence of POMDPs]
    \label{def:behaviouralEquivalencePOMDPs}
    We apply~\cref{def:behaviouralEquivalence} for $\type = \pomdp$:
    \begin{equation}
        % https://q.uiver.app/#q=WzAsNCxbMCwwLCJTIl0sWzAsMiwiTyBcXHRpbWVzIFNeSSJdLFsyLDAsIlQiXSxbMiwyLCJPIFxcdGltZXMgVF5JIl0sWzAsMSwiZiIsMl0sWzAsMiwiXFxwaGkiXSxbMiwzLCJnIl0sWzEsMywiXFx0ZXh0e2lkfV9PIFxcdGltZXMgXFxwaGleSSIsMl1d
        \begin{tikzcd}
            \States &&& \StatesFinal &&& \StatesPrime \\
            \\
            {\Dist(\Observations) \times \Dist(\States)^\Actions} &&& {\Dist(\Observations) \times \Dist(\StatesFinal)^\Actions} &&& {\Dist(\Observations) \times \Dist(\StatesPrime)^\Actions}
            \arrow["\transitionCoalg_\pomdp^\States"', from=1-1, to=3-1]
            \arrow["\beh_\States", from=1-1, to=1-4]
            \arrow["\transitionCoalg_\pomdp^\StatesFinal", from=1-4, to=3-4]
            \arrow["{\Dist(\id_\Observations) \times \Dist(\beh_\States)^\Actions}"', from=3-1, to=3-4]
            \arrow["\beh_\StatesPrime"', from=1-7, to=1-4]
            \arrow["\transitionCoalg_\pomdp^\StatesPrime", from=1-7, to=3-7]
            \arrow["{\Dist(\id_\Observations) \times \Dist(\beh_\StatesPrime)^\Actions}", from=3-7, to=3-4]
        \end{tikzcd}
    \end{equation}
\end{definition}

This corresponds to the following conditions (for more details see~\cref{apdx:concreteBehaviouralEquivalence}), where for any $\actions \in \Actions$ and $\statesFinal \in \StatesFinal$ we have:
\begin{align}
    \label{cond:behaviouralEquivalencePOMDPs}
    \dist(\observations \mid \states) & = \dist(\observations \mid \statesPrime) & \text{(condition 1)} \nonumber \\
    \sum_{\tilde{\states} \in \beh_\States^{-1}(\statesFinal)} \dist (\tilde{\states} \mid \states, \actions) & = \sum_{\tilde{\states}' \in \beh_\StatesPrime^{-1}(\statesFinal)} \dist (\tilde{\states}' \mid \statesPrime, \actions) & \text{(condition 2)}
\end{align}

This definition states that, given the same actions (by assumption), states $\states$ and $\statesPrime$ are behaviourally equivalent only if they emit the same probabilistic observations (by condition 1), while creating equivalence classes of indistinguishable ground truth states by considering their probabilistic transitions (by condition 2).
We believe this is too strict to properly describe predictive processing in all of its facets, as this requires probabilistic transitions of (equivalence classes of) ground truth states of the generative process, $\states \in \States$, to be equal to probabilistic transitions of (equivalence classes of) ground truth states of the generative model, $\statesPrime \in \StatesPrime$, while one of the main points of action-oriented generative models is that they don't need to recapitulate the veridical, ground truth structure of the environment~\cite{clarkRadicalPredictiveProcessing2015, clarkSurfingUncertaintyPrediction2015, baltieriActiveInferenceImplementation2017, baltieriGenerativeModelsParsimonious2019, baltieriPIDControlProcess2019, tschantzLearningActionorientedModels2020, mannellaActiveInferenceWhiskers2021}.

More generally, a definition of behavioural equivalence between probabilistic processes is notoriously non-trivial, and the formulation provided above is not the only possible choice (see e.g.~\cite{bonchiDistributionBisimilarityPower2021} for a review of this and other possible choices).
For probabilistic processes, it is in fact often desirable to focus on probabilistic properties rather than on characteristics of sampled trajectories from ground truth states as in~\cref{def:behaviouralEquivalencePOMDPs}.
We argue that this is also the case for predictive processing, which is based on the minimisation of the difference between distributions encoded by the generative model and generative process, rather than minimising the difference between trajectories sampled from them.

%
% -----------------------------------------------------------------------------------------

\subsubsection{Comparing belief MDPs}
Next, we will adapt the definition of \textbf{belief bisimulation equivalence}~\cite{castroEquivalenceRelationsFully2009, jansenBeliefBisimulationHidden2012}, originally restricted to a single system (hence the term ``equivalence'' (see~\cref{def:bisimulationEquivalence}), to work between different processes. In other words, we will define a \textbf{belief bisimulation}.
This definition corresponds to a standard bisimulation, that is, a span of coalgebras (see~\cref{def:bisimulation}) between coalgebras encoding \emph{beliefs}, in a Bayesian sense, as we shall see below, of the original processes.
This means that there is a corresponding notion of \textbf{belief behavioural equivalence} (a corelation or more generally a cospan of coalgebras, see~\cref{def:behaviouralEquivalence}), which once again is implied and implies that of bisimulation by working with well behaved functors and $\catset$ as the base category~\cite{mossFinalCoalgebrasFunctors2006}.

To apply the definition of belief behavioural equivalence, we start from a description of \emph{belief MDPs}~\cite{kaelblingPlanningActingPartially1998} associated to, or rather induced by POMDPs.
These are related to the separation principle of control~\cite{astromIntroductionStochasticControl1970} (see also~\cite{subramanianApproximateInformationState2022} for a review of related ideas).
Belief MDPs have previously been formulated in a coalgebraic context in~\cite{virgoUnifilarMachinesAdjoint2023, mcgregorFormalisingIntentionalStance2025a}, although they are not explicitly presented in terms of MDPs in those works.

\begin{definition}[Belief MDP]
    \label{def:BeliefMDP}
    A belief MDP induced by a POMDP $(\States, \Actions, \TransitionMap, \Observations, \ObservationMap)$ is an MDP $(\BeliefsFiltering, \Actions, \TransitionMap_\BeliefsFiltering)$ where:
    \begin{itemize}
        \item $\BeliefsFiltering$ is the space of belief states, sufficient statistics of histories 
        $\HistoriesFiltering \coloneqq (\Observations \times \Actions)^\ast \times \Observations$, given by $\beliefsFiltering: \HistoriesFiltering \to \Prob(\States)$,
        \item $\Actions$ is the space of actions and coincides with the one from the original POMDP,
        \item $\TransitionMap_\BeliefsFiltering : \BeliefsFiltering \times \Actions \to \Prob(\BeliefsFiltering)$ is the belief transitions dynamics, defined for $\beliefsFiltering_{t}, \beliefsFiltering_{t+1} \in \BeliefsFiltering$ and $\actions_{t} \in \Actions$ as
        \begin{align}
            \label{eqn:beliefTransitions}
            \TransitionMap_\BeliefsFiltering(\beliefsFiltering_{t}, \actions_{t}) &= \prob(\beliefsFiltering_{t+1} \mid \beliefsFiltering_{t}, \actions_{t}) \nonumber \\
            & = \sum_{\substack{\observations_{t} \in \Observations}
            } \prob(\beliefsFiltering_{t+1} \mid \beliefsFiltering_{t}, \observations_{t+1}, \actions_{t}) \prob(\observations_{t+1} \mid \beliefsFiltering_{t}, \actions_{t}),
        \end{align}
        where 
        \begin{equation}
            \prob(\beliefsFiltering_{t+1} \mid \beliefsFiltering_{t}, \observations_{t+1}, \actions_{t}) =
            \left\{
            \begin{aligned}
                & 1 \quad \text{if } \beliefUpdateFiltering(\beliefsFiltering_{t}, \observations_{t+1}, \actions_{t}) = \beliefsFiltering_{t+1}, \\
                & 0 \quad \text{otherwise},
            \end{aligned}
            \right.
        \end{equation}
        for $\beliefUpdateFiltering : \BeliefsFiltering \times \Observations \times \Actions \to \BeliefsFiltering$ defined by standard Bayesian filtering updates of beliefs $\beliefsFiltering_{t} = \prob(\states_{t} \mid \historiesFiltering_{t})$ for $\historiesFiltering_t \in \HistoriesFiltering_t$, see~\cite{kaelblingPlanningActingPartially1998, rosasAIVatFundamental2025} and~\cref{apdx:bayesianFiltering}.
    \end{itemize}
\end{definition}

In a belief MDP, beliefs, probability distributions over the possible states, serve as the states of a standard MDP.%
Applying this construction to both the generative model and generative process, given as POMDPs or probabilistic Moore machines in coalgebraic form (see~\cref{def:categoryPOMDP}), produces two belief MDPs, describing the associated probability distributions and their transitions (obtained by currying, see~\cref{ex:MooreMachines}):
\begin{align}
    \TransitionMap_\BeliefsFiltering : \BeliefsFiltering \times \Actions \to \Prob(\BeliefsFiltering) \quad \leftrightarrow \quad \transitionCoalg_\mdp^\BeliefsFiltering & : \BeliefsFiltering \to \Dist(\BeliefsFiltering)^\Actions \quad \text{(belief generative process)} \nonumber \\
    \TransitionMap_\BeliefsFilteringPrime : \BeliefsFilteringPrime \times \Actions \to \Prob(\BeliefsFilteringPrime) \quad \leftrightarrow \quad \transitionCoalg_\mdp^\BeliefsFilteringPrime & : \BeliefsFilteringPrime \to \Dist(\BeliefsFilteringPrime)^\Actions \quad \text{(belief generative model)}.
\end{align}

Using these, a belief behavioural equivalence between them can be defined as follows:

\begin{definition}[Belief behavioural equivalence of belief MDPs]
    \label{def:behaviouralEquivalencePOMDPsBeliefs}
    This is a direct application of~\cref{def:behaviouralEquivalence} (once again, without visualising $R$ for simplicity) for $\type = \mdp$:
    \begin{equation}
        % https://q.uiver.app/#q=WzAsNCxbMCwwLCJTIl0sWzAsMiwiTyBcXHRpbWVzIFNeSSJdLFsyLDAsIlQiXSxbMiwyLCJPIFxcdGltZXMgVF5JIl0sWzAsMSwiZiIsMl0sWzAsMiwiXFxwaGkiXSxbMiwzLCJnIl0sWzEsMywiXFx0ZXh0e2lkfV9PIFxcdGltZXMgXFxwaGleSSIsMl1d
        \begin{tikzcd}
            \BeliefsFiltering &&& \StatesFinal &&& \BeliefsFilteringPrime \\
            \\
            {\Dist(\BeliefsFiltering)^\Actions} &&& {\Dist(\StatesFinal)^\Actions} &&& {\Dist(\BeliefsFilteringPrime)^\Actions}
            \arrow["\transitionCoalg_\mdp^\BeliefsFiltering"', from=1-1, to=3-1]
            \arrow["\beh_\BeliefsFiltering", from=1-1, to=1-4]
            \arrow["\transitionCoalg_\mdp^\StatesFinal", from=1-4, to=3-4]
            \arrow["{\Dist(\beh_\BeliefsFiltering)^\Actions}"', from=3-1, to=3-4]
            \arrow["\beh_\BeliefsFilteringPrime"', from=1-7, to=1-4]
            \arrow["\transitionCoalg_\mdp^\BeliefsFilteringPrime", from=1-7, to=3-7]
            \arrow["{\Dist(\beh_\BeliefsFilteringPrime)^\Actions}", from=3-7, to=3-4]
        \end{tikzcd}
    \end{equation}
    This coincides with the following definition, i.e. condition 2 in~\cref{cond:behaviouralEquivalencePOMDPs} with a different state space: beliefs on states, rather than states, and without observations due to the Bayesian construction  in~\cref{def:BeliefMDP} (this can be obtained as in~\cref{apdx:concreteBehaviouralEquivalence}, with trivial observations, i.e. $\Observations = 1$), where, for any $\actions \in \Actions$ and $\statesFinal \in \StatesFinal$, we have:
    \begin{align}
        \label{cond:behaviouralEquivalencePOMDPsBelief}
        \sum_{\tilde{\beliefsFiltering} \in \beh_\BeliefsFiltering^{-1}(\statesFinal)} \dist(\tilde{\beliefsFiltering} \mid \beliefsFiltering, \actions) & = \sum_{\tilde{\beliefsFiltering}' \in \beh_\BeliefsFilteringPrime^{-1}(\statesFinal)} \dist(\tilde{\beliefsFiltering}' \mid \beliefsFilteringPrime, \actions).
    \end{align}
\end{definition}

This condition implies that beliefs $\beliefsFiltering$ and $\beliefsFilteringPrime$ are behaviourally equivalent if the distributions over the respective next states, belonging to the same equivalence class represented by \(\statesFinal\), are equal given the same actions, but does not make any statement about observations.
This is a consequence of the definition of belief MDPs (\cref{def:BeliefMDP}), in which observations are marginalised in the definition of belief updates (see~\cref{eqn:beliefTransitions}).
Such a condition could be relevant in situations where we require the beliefs of two processes to be equivalent: their beliefs evolve in the same way, while the exact implementations of these beliefs are not important.
It is however problematic for another, crucial reason: the final coalgebra of the category of MDPs is trivial, i.e. $\StatesFinal$ is the one-element set. \footnote{The authors would like to thank Nathaniel Virgo for pointing this out, see also~\cite{devinkBisimulationProbabilisticTransition1999, sokolovaProbabilisticSystemsCoalgebraically2011} for related results.}
This means that belief MDPs are actually \emph{not observable} in the coalgebraic sense~\cite{jacobsIntroductionCoalgebraMathematics2017}, since there is no non-trivial monomorphism (in our setup, an injective map) into the final coalgebra (since it has only one element).
We thus turn to another approach to describe similarity between the generative model and generative process.

%
% -----------------------------------------------------------------------------------------

\subsubsection{Comparing lifted POMDPs}
In our final attempt to find a behavioural notion of similarity between the generative model and generative process, we turn to a generalisation of \textbf{distribution bisimulation equivalence}~\cite{crafaSpectrumBehavioralRelations2011, silvaGeneralizingDeterminizationAutomata2013, fengWhenEquivalenceBisimulation2014, hermannsProbabilisticBisimulationNaturally2014} between processes, i.e. \textbf{distribution bisimulation}.
Under the assumptions of this work, this also yields a notion of \textbf{distribution behavioural equivalence}.

Note that work on this type of equivalence is often grouped with what we have described as belief bisimulation and behavioural equivalence (see, e.g.~\cite{bonchiDistributionBisimilarityPower2021}).
However, we wish to emphasise that, while both are behavioural equivalences on probability distributions, the compared distributions have a markedly different semantics. 
In belief equivalence, the distributions are constructed as Bayesian beliefs on hidden states of a given process. 
In distribution bisimulations, by contrast, they correspond to probability distributions on hidden states that are not necessarily updated using Bayes~\cite{silvaGeneralizingDeterminizationAutomata2013}.
While the \emph{determinisation} of a POMDP is another POMDP, a \emph{belification} of a POMDP is a (belief) MDP.
We further note that these are also related to the \emph{unifilarisation} described in~\cite{virgoInterpretingDynamicalSystems2021}, but do not explore here the details.

We start by recalling the generalised determinisation construction of~\cite{silvaGeneralizingDeterminizationAutomata2013}, focusing only on POMDPs.
\begin{definition}[Generalised determinisation of POMDPs]
    Given a POMDP as a coalgebra of type $(\States, \transitionCoalg_\pomdp^\States : \States \to \Dist(\Observations) \times \Dist(\States)^\Actions)$, the generalised determinisation of $(\States, \transitionCoalg_\pomdp^\States)$ is the lifted coalgebra of type $(\Dist(\States), {\transitionCoalg_\pomdp^\States}^\sharp : \Dist(\States) \to \Dist(\Observations) \times \Dist(\States)^\Actions)$ such that the following diagram commute:
    \begin{equation}
        \begin{tikzcd}
            S && {\Dist(\States)} \\
            \\
            {\Dist(\Observations) \times \Dist(\States)^\Actions}
            \arrow["{\eta_\States}", from=1-1, to=1-3]
            \arrow["{\langle \transitionCoalgOut_\pomdp^\States, \: \transitionCoalgTr_\pomdp^\States \rangle}"', from=1-1, to=3-1]
            \arrow["{\langle {\transitionCoalgOut_\pomdp^\States}^\sharp, \: {\transitionCoalgTr_\pomdp^\States}^\sharp \rangle}", from=1-3, to=3-1]
        \end{tikzcd}
    \end{equation}
    or in other words, such that:
    \begin{align}
        {\transitionCoalgOut_\pomdp^\States} & = \eta_\States \comp {\transitionCoalgOut_\pomdp^\States}^\sharp \quad \text{ and} \nonumber \\
        {\transitionCoalgTr_\pomdp^\States} & = \eta_\States \comp {\transitionCoalgTr_\pomdp^\States}^\sharp,
    \end{align}
    where $\eta_\States: \States \to \Dist(\States)$ is a map\footnote{For readers familiar with it, this is the unit of distribution monad, since $\Dist$ is not only a functor but a full fledged monad~\cite{maclaneCategoriesWorkingMathematician1978}.} that, given $\States$, returns the (Kronecker) delta distribution of $\States$, $\delta_\States$, and the lifted maps\footnote{For readers familiar with it, this is just a Kleisli extension, i.e. given the multiplication of the distribution monad $\mu_X$ and a morphism $g$, we have $g^\sharp = \Dist(g) \comp \mu_X$.} $\langle {\transitionCoalgOut_\pomdp^\States}^\sharp, {\transitionCoalgTr_\pomdp^\States}^\sharp \rangle$ are given, for any $\beliefsPredicting \in \Dist(\States), \actions \in \Actions$, and $\states, \tilde{\states} \in \States$ by: 
    \begin{align}
        {\transitionCoalgOut_\pomdp^\States}^\sharp (\beliefsPredicting)(\observations) 
        & = \sum_{\states \in \States} \beliefsPredicting(\states) \ {\transitionCoalgOut_\pomdp^\States}(\states)(\observations) \quad \text{ and} \nonumber \\
        {\transitionCoalgTr_\pomdp^\States}^\sharp (\beliefsPredicting)(\actions)(\tilde{\states}) 
        & = \sum_{\states \in \States} \beliefsPredicting(\states) \ {\transitionCoalgTr_\pomdp^\States}(\states)(\actions)(\tilde{\states})
    \end{align}
    or in a more traditional notation:
    \begin{align}
        \label{eqn:liftedTransitionsObservations}
        \dist(\observations \mid \beliefsPredicting) & = \sum_{\states \in \States} \beliefsPredicting(\states) \ \dist(\observations \mid \states) \quad \text{ and} \nonumber \\
        \dist(\tilde{\states} \mid \beliefsPredicting, \actions) & = \sum_{\states \in \States} \beliefsPredicting(\states) \ \dist(\tilde{\states} \mid \states, \actions).
    \end{align}
\end{definition}

Determinisation allows us to work with a coalgebraic structure in which the transition map, ${\langle {\transitionCoalgOut_\pomdp^\States}^\sharp, \: {\transitionCoalgTr_\pomdp^\States}^\sharp \rangle}$, is from a space of distributions over states, $\Dist(\States)$, to the space of distributions over the next state and observations, instead of being from the state space $\States$ itself.
As we will see next, in this way, we can think of the transition dynamics in the POMDP as a transition from one state distribution to another.
Further, based on this notion of determinisation, we can compare POMDPs in terms of probability distributions over states and observations, without inducing a belief MDP that marginalises observations, see the belief transition dynamics in~\cref{def:BeliefMDP}.
To see this, we proceed to define the notion of lifted POMDP.

\begin{definition}[Lifted POMDP]
    \label{def:LiftedPOMDP}
    A lifted POMDP induced by a POMDP $(\States, \Actions, \TransitionMap, \Observations, \ObservationMap)$ via determinisation~\cite{silvaGeneralizingDeterminizationAutomata2013} is a POMDP $(\BeliefsPredicting, \Actions, \TransitionMap_\BeliefsPredicting, \Observations, \ObservationMap_\BeliefsPredicting)$ where:
    \begin{itemize}
        \item $\BeliefsPredicting$ is the space of belief states $\Dist(\States)$\footnote{NB: $\BeliefsPredicting \ne \BeliefsFiltering$ in general, however their relation won't be explored further here.},
        \item $\Actions$ is the space of actions and coincides with the one from the original POMDP,
        \item $\TransitionMap_\BeliefsPredicting : \BeliefsPredicting \times \Actions \to \BeliefsPredicting$ is the belief transitions dynamics, defined for $\beliefsPredicting_{t}, \beliefsPredicting_{t+1} \in \BeliefsPredicting$ and $\actions_{t} \in \Actions$ as:
        \begin{align}
            \TransitionMap_\BeliefsPredicting(\beliefsPredicting_{t}, \actions_{t}) &= \beliefsPredicting_{t+1} = \beliefsPredicting(\states_{t+1}) = \prob(\states_{t+1} \mid \beliefsPredicting_{t}, \actions_t)
            \nonumber \\
            &
            = \sum_{\states_{t} \in \States} \beliefsPredicting(\states_{t}) \ \prob(\states_{t+1} \mid \states_{t}, \actions_t)
        \end{align}
        and correspond to the lifted transition map in~\cref{eqn:liftedTransitionsObservations},
        \item $\ObservationMap_\BeliefsPredicting : \BeliefsPredicting \to \Prob(\Observations)$ is the belief observation map defined for $\beliefsPredicting_{t} \in \BeliefsPredicting$, $\actions_{t} \in \Actions$ and $\observations_t \in \Observations$ as:
        \begin{align}
            \ObservationMap_\BeliefsPredicting(\beliefsPredicting_{t}) & = \prob(\observations_{t} \mid \beliefsPredicting_t) \nonumber \\
            & = \sum_{\states_{t} \in \States} \beliefsPredicting_{t}(\states_{t}) \ \prob(\observations_{t} \mid \states_{t}),
        \end{align}
        and correspond to the lifted observation map in~\cref{eqn:liftedTransitionsObservations}.
    \end{itemize}
    We note that this is a rather special kind of POMDP, one in which state transitions are deterministic.
    In some sense, this is a different generalisation of deterministic Moore machines: probabilistic Moore machines in~\cref{def:categoryPOMDP} make both transition and observation maps stochastic, while here only the observation map is stochastic.
\end{definition}

Applying the generalised determinisation of~\cite{silvaGeneralizingDeterminizationAutomata2013} to the generative process and generative model as coalgebras yields the following, respectively (by currying, see~\cref{ex:MooreMachines}):
\begin{align}
    \langle \transitionCoalgOut_\pomdp^\BeliefsPredicting, \transitionCoalgTr_\pomdp^\BeliefsPredicting \rangle & : \BeliefsPredicting \to \Dist(\Observations) \times \BeliefsPredicting^\Actions \quad & \text{(lifted generative process)} \nonumber \\
    \langle \transitionCoalgOut_\pomdp^\BeliefsPredictingPrime, \transitionCoalgTr_\pomdp^\BeliefsPredictingPrime \rangle & : \BeliefsPredictingPrime \to \Dist(\Observations) \times \BeliefsPredictingPrime^\Actions \quad & \text{(lifted generative model)}.
\end{align}

Using these, a distribution behavioural equivalence between them amounts to the following:

\begin{definition}[Distribution behavioural equivalence of lifted POMDPs]
    \label{def:behaviouralEquivalencePOMDPsLifted}
    This is a direct application of~\cref{def:behaviouralEquivalence} for $\type = \pomdp$, with POMDPs given as lifted POMDPs from~\cref{def:LiftedPOMDP} (once again, without visualising $R$ for simplicity):
    \begin{equation}
        \begin{tikzcd}
            \BeliefsPredicting &&& \StatesFinal &&& \BeliefsPredictingPrime \\
            \\
            {\Dist(\Observations) \times \BeliefsPredicting^\Actions} &&& {\Dist(\Observations) \times \StatesFinal^\Actions} &&& {\Dist(\Observations) \times \BeliefsPredictingPrime^\Actions}
            \arrow["{\langle \transitionCoalgOut_\pomdp^\BeliefsPredicting, \transitionCoalgTr_\pomdp^\BeliefsPredicting \rangle}"', from=1-1, to=3-1]
            \arrow["\beh_\BeliefsPredicting", from=1-1, to=1-4]
            \arrow["{\langle \transitionCoalgOut_\pomdp^\StatesFinal, \transitionCoalgTr_\pomdp^\StatesFinal \rangle}", from=1-4, to=3-4]
            \arrow["{\Dist(\id_\Observations) \times (\beh_\BeliefsPredicting)^\Actions}"', from=3-1, to=3-4]
            \arrow["\beh_\BeliefsPredictingPrime"', from=1-7, to=1-4]
            \arrow["{\langle \transitionCoalgOut_\pomdp^\BeliefsPredictingPrime, \transitionCoalgTr_\pomdp^\BeliefsPredictingPrime \rangle}", from=1-7, to=3-7]
            \arrow["{\Dist(\id_\Observations) \times (\beh_\BeliefsPredictingPrime)^\Actions}", from=3-7, to=3-4]
        \end{tikzcd}
    \end{equation}
    which corresponds to the following conditions (see again~\cref{apdx:concreteBehaviouralEquivalence}, considering that transitions are deterministic and hence delta distributions), where for any $\actions \in \Actions, \beliefsPredicting \in \BeliefsPredicting$ and $\beliefsPredictingPrime \in \BeliefsPredictingPrime$ we have:
    \begin{align}
        \label{cond:behaviouralEquivalencePOMDPsLifted}
        \dist(\observations \mid \beliefsPredicting) & = \dist(\observations \mid \beliefsPredictingPrime) & \text{(condition 1)} \nonumber \\
        \beh_\BeliefsPredicting \big( \transitionCoalgTr_\pomdp^\BeliefsPredicting(\beliefsPredicting)(\actions) \big) &= \beh_{\BeliefsPredictingPrime} \big( \transitionCoalgTr_\pomdp^\BeliefsPredictingPrime(\beliefsPredictingPrime)(\actions) ) & \text{(condition 2)}
    \end{align}
\end{definition}

Condition 1 states that two beliefs, $\beliefsPredicting$ and $\beliefsPredictingPrime$, are behaviourally equivalent only if they produce the same expected probability of observations (see~\cref{eqn:liftedTransitionsObservations}).
In other words, provided that condition 2 also holds, $\beliefsPredicting$ and $\beliefsPredictingPrime$ are equivalent if their distributions on states ``average out'' to the same observations.
They may represent different internal (i.e. state) information, yet their expected observational consequences are identical.

The second condition is recursive in nature, as is typical for bisimulations of deterministic systems~\cite{sangiorgiAdvancedTopicsBisimulation2011}. 
It states that, for two beliefs $\beliefsPredicting$ and $\beliefsPredictingPrime$ to be behaviourally equivalent, their predicted future beliefs must also be equivalent for any given action.
This implies that two beliefs are indistinguishable if they lead to the same beliefs (predictions) about the next state of the world, that is, equivalence is tested on the belief that results from pure dynamical prediction, without taking into account new evidence, i.e. observations, which are instead part of condition 1.
This contrasts with belief MDPs, where these two conditions are combined into a condition on Bayesian updates (see~\cref{cond:behaviouralEquivalencePOMDPsBelief}).

%
%
%
% -----------------------------------------------------------------------------------------
% -----------------------------------------------------------------------------------------
% -----------------------------------------------------------------------------------------

\section{Discussion}
\begin{figure}[!ht]
    \centering
    \includegraphics[width=0.7\linewidth]{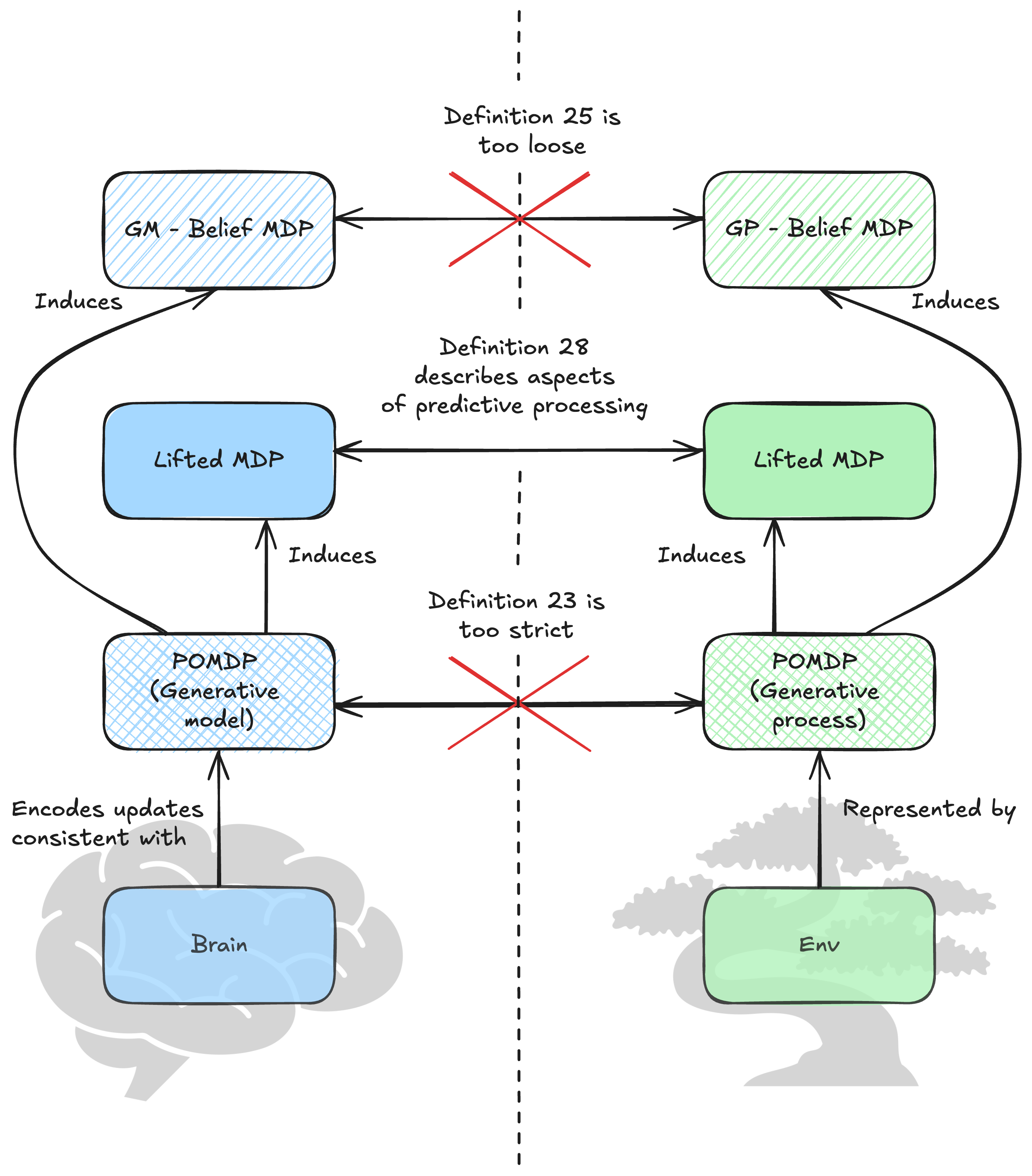}
    \caption{The three possible descriptions of behavioural equivalence introduced in this work.}
    \label{fig:bisimulation}
\end{figure}

In~\cref{sec:behaviouralEquivalencesPP}, we introduced three distinct notions of behavioural equivalence building on~\cref{def:behaviouralEquivalence}: behavioural equivalence of POMDPs (\cref{def:behaviouralEquivalencePOMDPs}), belief behavioural equivalence of belief POMDPs (\cref{def:behaviouralEquivalencePOMDPsBeliefs}), and distribution behavioural equivalence of lifted POMDPs (\cref{def:behaviouralEquivalencePOMDPsLifted}).
Translating the conditions of each definition into a more familiar form gives us some background on their implications and relations to predictive processing.
We summarise this high level account pictorially in~\cref{fig:bisimulation}.

\cref{cond:behaviouralEquivalencePOMDPs} suggests that~\cref{def:behaviouralEquivalencePOMDPs} may be too strict, as it requires observations for particular ground truth states to be equal.
While states can be coarse grained based on transition dynamics, the condition on observations seems too strong.
In contrast, ~\cref{def:behaviouralEquivalencePOMDPsBeliefs} requires only Bayesian beliefs to be equal, under the assumption that they can be coarse grained if their transitions allow for it (see~\cref{cond:behaviouralEquivalencePOMDPsBelief}).
From a coalgebraic perspective, however, this condition is too loose: based on the definition of behavioural equivalence given in~\cref{def:behaviouralEquivalence}, this condition must be satisfied for elements of the final coalgebra of the category of MDPs (which contain belief MDPs).
Such final coalgebra is however trivial, a one element set, which implies that \emph{all} belief MDPs can be said to be behaviourally equivalent under this account, because their states are never exposed to an external observer (see ~\cref{sec:limitations} for further discussion).
The third proposal seemingly hits a sweet spot, and we believe it can be used to establish some of the relevant connections between coalgebras, their language to compare systems' behaviours and predictive processing.

\cref{def:behaviouralEquivalencePOMDPsLifted} translates some of the core principles of predictive processing and active inference in a coalgebraic language, clarifying the goal, the mechanism, and the nature of an agent's generative model.
Condition 1 in~\cref{def:behaviouralEquivalencePOMDPsLifted} defines the goal of predictive processing in terms of prediction error minimisation, which is expressed as the idea that the agent's model must generate the same sensory predictions as the environment.
If we take $\beliefsPredicting \in \BeliefsPredicting$ to represent probability distributions over states of the environment encoded by the generative process, Condition 1 demands that the agent's beliefs $\beliefsPredictingPrime \in \BeliefsPredictingPrime$, which, it should be noted, are distinct from those in a belief MDP, encoded by the implicit generative model produce the exact same probability distribution over observations $\observations \in \Observations$ given by the environment.
In active inference, an agent that has minimised its free energy to the greatest possible extent is one that has successfully tuned its model to satisfy this condition.

Furthermore, an active inference agent never has direct access to the true states of the world $\states \in \States$, but only to its own probabilistic beliefs about the world.
Distribution behavioural equivalence operates at the level of these beliefs, providing a mechanism to compare an agent's beliefs, $\varprob{\States_{1:\ntime}, \policy, \obsmap, \transmap}$, which correspond ($D_\text{KL} = 0$ in the limit) to the exact posterior $\condprob{\States_{1:\ntime}, \policy, \obsmap, \transmap}{\obsvar_{1:\ntime}}$ for agents performing exact inference (see~\cref{eqn:aif-kl-fe}), with the environment's ground truth probabilistic state, rather than comparing an agent's internal states with the world's true hidden states, cf.~\cref{def:behaviouralEquivalencePOMDPs}.

Thus, behavioural equivalence well captures what we believe to be the true nature of active inference, proposing that the brain's implicit generative models need only be action-oriented, enabling an agent to fulfil its goals but not necessarily proving to be perfect and exhaustive copies of the generative processes of the world.
In this light, a generative model, with state space $\StatesPrime$, can be vastly simpler than the environment's generative process with state space $\States$.
As long as the two systems satisfy the conditions in~\cref{cond:behaviouralEquivalencePOMDPsLifted}, making the same predictions (condition 1) and evolving their beliefs in indistinguishable ways (condition 2), they are functionally equivalent.
This framework thus demonstrates \emph{when} a simpler model can be ``good enough'', by formalizing what it means to be observationally indistinguishable.

\subsection{Related work}
This perspective is based on a coalgebraic formulation of dynamical systems, describing both a generative process of the environment and an implicit generative model encoded by brain states.
The application of the coalgebraic language to MDPs is not new, see, for instance~\cite{feysLongTermValuesMarkov2018}, which defines a category of MDPs that includes rewards (cf.~\cref{def:categoryMDP}) and~\cite{virgoUnifilarMachinesAdjoint2023, mcgregorFormalisingIntentionalStance2025a}.
Coalgebras have also been applied in the context of predictive processing (see for instance~\cite{smitheCompositionalActiveInference2022a, smitheOpenDynamicalSystems2023}).
However, these works make no explicit reference to the structural reading of active inference and predictive processing proposed here, highlighting the generative model and process as two distinct entities, clarifying their role within the overall framework, and showing how to use maps of systems, particularly bisimulations, to understand the purported synchronisation of brain and environment (although~\cite{smitheOpenDynamicalSystems2023} defines a notion of (quasi-)bisimulations to get to a general definition of Bayesian inversions in their setup).

With the use of a coalgebraic framework, we have also seen how immediate it is to switch between seemingly different kinds of dynamical systems: deterministic (open) dynamical systems, or Moore machines, possibilistic Moore machines and probabilistic Moore machines are, in fact all examples of ``structured Moore machines'' as defined in~\cite{silvaGeneralizingDeterminizationAutomata2013}.
That is to say, their transition types, are the same: given an input, a system combines the current state and the input to obtain the next (one/possible/distribution of) state(s) and produce (one/possible/a distribution of) output(s). 
What changes is simply the content in the brackets, differentiating the kind of branching a system can have, e.g. deterministic, possibilistic or probabilistic.
It is thus, to some extent, not surprising to see important works such as~\cite{isomuraTripleEquivalenceEmergence2025} highlighting the similarities between formulations of predictive processing and classical automata. However, we should stress that, following~\cite{silvaGeneralizingDeterminizationAutomata2013}, one cannot simply expect to use another kind of structured Moore machine to obtain Turing machines, so the precise connection to~\cite{isomuraTripleEquivalenceEmergence2025}, if it exists at all, will be investigated in future work.

We believe that our work based on behavioural equivalence of generative model and generative process is also related to the ``interpretation map'' approach in~\cite{virgoInterpretingDynamicalSystems2021, biehlInterpretingSystemsSolving2022}, however we do not explore possible connections here.
Our definition of lifted POMDP, see~\cref{def:LiftedPOMDP}, employs a notion of belief states that is also potentially related to the definition of predictive beliefs in~\cite{rosasAIVatFundamental2025}.
This could then explain its relation to the beliefs in a belief MDP (\cref{def:BeliefMDP}), which would be their postdictive counterpart, however, this remains at present speculative.~\footnote{Roughly: predictive beliefs are sufficient statistics built from a Bayesian prediction step, a pushforward of probabilities along the dynamics of the system before a new observation is collected, while postdictive beliefs are sufficient statistics generated after the observation is collected from a Bayesian update step.}

%
%
% -----------------------------------------------------------------------------------------
% -----------------------------------------------------------------------------------------

\subsection{Current limitations and future work}
\label{sec:limitations}

Currently, we do not handle continuous probabilities (cf. Giry functor on measurable spaces~\cite{jacobsIntroductionCoalgebraMathematics2017}).
While discrete probabilities are a helpful simplification and are also useful for different parts of active inference and predictive processing~\cite{dacostaActiveInferenceDiscrete2020}, they do not cover the whole framework, which also involves large parts using continuous probabilities~\cite{fristonFreeEnergyPrinciple2019, fristonFreeEnergyPrinciple2023}.
One of the main reasons to leave a similar treatment for continuous probabilities to future work is the fact that definitions of bisimulation for continuous probabilities are more difficult to handle and require more careful considerations~\cite{jacobsIntroductionCoalgebraMathematics2017, hermidaBisimulationLogicalRelation2022, moroniClassificationBisimilaritiesGeneral2024}.

We also do not currently discuss several other computational components of active inference, including learning, policy selection, and action, which are a central part of this framework (for a review, see~\cite{dacostaActiveInferenceDiscrete2020}) but for which we lack a coalgebraic counterpart.
The algorithmic part of active inference is also currently not discussed, as our focus is on ``fixed points'', that is, the behavioural equivalence of generative model and generative process does not really include phases like the transients leading to synchronisation.
This is in part because the current work is focused on laying down the foundations for a coalgebraic treatment of some aspects of predictive processing and active inference, and in part because we are not currently aware of algorithmic implementations compatible with behavioural equivalence that don't rely on an underlying bisimulation \emph{equivalence}.
This is perhaps not a major limitation since, bisimulations between two coalgebras (which are in a one-to-one correspondence to behavioural equivalences under the right conditions) are equivalent to bisimulation equivalences\footnote{Following standard results (see for instance~\cite[proposition 3.44 in the October 8$^\text{th}$ 2024 version]{moroniClassificationBisimilaritiesGeneral2024}), given two coalgebras $\mathcal{S} = (S, f^S), \mathcal{S}' = (S', f^{S'})$, a cospan between $\mathcal{S}$ and $\mathcal{S}'$ (``external'' behavioural equivalence) is equivalent to a cospan between $\mathcal{S} + \mathcal{S}'$, the disjoint union (or coproduct) of the two original coalgebras, and $\mathcal{S} + \mathcal{S}'$, i.e. itself (``internal'' behavioural equivalence).}, for which various implementations of learning algorithms (i.e. exhibiting transients) exist in machine learning and reinforcement learning where bisimulation equivalences are a central research topic~\cite{zhangLearningCausalState2021, zhangLearningInvariantRepresentations2021, rosasAIVatFundamental2025}. 
In these fields, they are used to describe compressions (see their relation to models in~\cref{def:model}) of state or state(-action) spaces of a (PO)MDP, often using approximation such as ``bisimulation metrics''~\cite{fernsBisimulationMetricsAre2014}.

Finally, our definition of a category of MDPs (\cref{def:categoryMDP}) can be potentially improved: while it does have a final coalgebra, this is trivial (i.e. the one-element set), which renders all non-trivial coalgebras in this category non observable in a coalgebraic sense, that is, their maps into the final coalgebra are not monomorphisms.
This is perhaps counter-intuitive from the perspective of reinforcement learning and active inference, where MDPs are traditionally seen as fully observable.
This observation points to a potential inconsistency that may need to be addressed.
Technically, we could change the definition of the category of MDPs so that these coalgebras produce an output, exposing their states directly, i.e. $\mdp(\States) = \States \times \Dist(\States)^\Actions$.
However, this would produce systems that expose all possible states $\States$ rather than the particular state at a specific time step.\footnote{The authors would like to thank Nathaniel Virgo for pointing this out.}
It would be a problem for the traditional definition of bisimulation. 
In that case, we would either need to assume that given coalgebras $(\States, \transitionCoalg^\States), (\StatesPrime, \transitionCoalg^\StatesPrime)$, their states are equal (because their exposes outputs would need to be equal, see~\cref{def:behaviouralEquivalencePOMDPs} where in place of $\Dist(\Observations)$ we would have $\States$), or abandon the standard notion of behavioural equivalence if their outputs are to remain distinct.
Alternatively, we could adopt the definition from~\cite{feysLongTermValuesMarkov2018}, with coalgebras of the form $\States \to \mathbb{R} \times \Dist(\States)^\Actions$, where $\mathbb{R}$ is used to represent rewards.
This category has a non-trivial final coalgebra because rewards are observable. 
However, since active inference and predictive processing do not usually rely on rewards, this would be outside the scope of this work.

%
%
%
% -----------------------------------------------------------------------------------------
% -----------------------------------------------------------------------------------------
% -----------------------------------------------------------------------------------------

\section{Conclusion}
In this work we captured important properties of agent-environment coupled systems where agents are thought to implement a process of free energy minimisation in predictive processing, using the language of coalgebras~\cite{ruttenUniversalCoalgebraTheory2000, jacobsIntroductionCoalgebraMathematics2017}.
More precisely, we provided a new and more formal perspective on the idea that an agent's implicit generative model need not in fact be isomorphic in a structural sense (cf. algebras vs. coalgebras~\cite{ruttenUniversalCoalgebraTheory2000, barbosaAlgebraicCoalgebraicStructures2005, venemaAlgebrasCoalgebras2007, ruttenMethodCoalgebraExercises2019}) to the generative process representing relevant parts of the environment~\cite{baltieriActiveInferenceImplementation2017, baltieriGenerativeModelsParsimonious2019, baltieriPIDControlProcess2019, tschantzLearningActionorientedModels2020, mannellaActiveInferenceWhiskers2021}.
Instead, we argued that what truly matters is the \emph{behaviour} of an agent's brain, or rather of the generative model as a POMDP that it implicitly encodes, which in coalgebraic terms corresponds to its outputs over time, i.e. predictions of observations over different modalities.
In particular, it matters that these predictions are compatible with the observations produced by the environment's generative process on a distribution level, rather than for ground truth states of the brain and the environment (\cref{def:behaviouralEquivalencePOMDPsLifted}), while remaining consistent with achieving the agent's overall goals.

\section*{Acknowledgments}
M.B. and T.N. were supported by JST FOREST Program (JPMJFR231V). T.N. was also supported by JSPS KAKENHI (grant numbers JP24H02172 and JP24H01559).
F.T. was supported by JST, Moonshot R\&D, Grant Number JPMJMS2012. 

\beginappendix

%
%
%
% -----------------------------------------------------------------------------------------
% -----------------------------------------------------------------------------------------
% -----------------------------------------------------------------------------------------

\section{Concrete behavioural equivalence for POMDPs}
\label{apdx:concreteBehaviouralEquivalence}

Given two coalgebras $(\States, \transitionCoalg_\pomdp^{\States})$ and $(\StatesPrime, \transitionCoalg_\pomdp^{\StatesPrime})$, two states $\states \in \States$ and $\statesPrime \in \StatesPrime$ are behaviourally equivalent if there exists a final $\type$-coalgebra $(\StatesFinal, \transitionCoalg_\pomdp^\StatesFinal = (\transitionCoalgTr_\pomdp^\StatesFinal (\states), \transitionCoalgOut_\pomdp^\StatesFinal (\states)))$ and a pair of $\type$-coalgebra morphisms $\beh_{\States}: (\States, \transitionCoalg_\pomdp^{\States}) \rightarrow(\StatesFinal, \transitionCoalg_\pomdp^\StatesFinal)$ and $\beh_{\StatesPrime}: (\StatesPrime, \transitionCoalg_\pomdp^\StatesPrime) \rightarrow(\StatesFinal, \transitionCoalg_\pomdp^\StatesFinal)$ such that $\beh_{\States}(\states) = \beh_\StatesPrime(\statesPrime)$.
This can be written in a more familiar form.
To see that, let $\statesFinal_0 = \beh_{\States} (\states) = \beh_\StatesPrime (\statesPrime)$.
The maps $\beh_{\States}$ and $\beh_\StatesPrime$ are $\type$-coalgebra morphisms if both:
\begin{align}
    \transitionCoalg_\pomdp^\StatesFinal \big(\beh_{\States}(\states)\big) (\actions)
    & = \big( \Dist(\id_\Observations) \times \Dist (\beh_{\States}) \big) \transitionCoalg_\pomdp^{\States}(\states) (\actions) \nonumber \\
    & = \big( \Dist(\id_\Observations) \times \Dist (\beh_{\States}) \big) \big\langle \transitionCoalgOut_\pomdp^\States (\states), \transitionCoalgTr_\pomdp^\States (\states) (\actions) \big\rangle \nonumber \\
    & = \big\langle \Dist(\id_\Observations) \transitionCoalgOut_\pomdp^\States (\states), \Dist (\beh_{\States}) \transitionCoalgTr_\pomdp^\States (\states) (\actions) \big\rangle \nonumber \\
    & = \big\langle \transitionCoalgOut_\pomdp^\States (\states), \Dist (\beh_{\States}) \transitionCoalgTr_\pomdp^\States (\states) (\actions) \big\rangle 
\end{align}
and
\begin{align}
    \transitionCoalg_\pomdp^\StatesFinal \big(\beh_\StatesPrime(\statesPrime)\big) (\actions)
    & = \big( \Dist(\id_\Observations) \times \Dist (\beh_\StatesPrime) \big) \transitionCoalg_\pomdp^\StatesPrime(\statesPrime) (\actions) \nonumber \\
    & = \big( \Dist(\id_\Observations) \times \Dist (\beh_\StatesPrime) \big) \big\langle \transitionCoalgOut_\pomdp^\States (\statesPrime), \transitionCoalgTr_\pomdp^\States (\statesPrime) (\actions) \big\rangle \nonumber \\
    & = \big\langle \Dist(\id_\Observations) \transitionCoalgOut_\pomdp^\States (\statesPrime), \Dist (\beh_\StatesPrime) \transitionCoalgTr_\pomdp^\States (\statesPrime) (\actions) \big\rangle \nonumber \\
    & = \big\langle \transitionCoalgOut_\pomdp^\States (\statesPrime), \Dist (\beh_\StatesPrime) \transitionCoalgTr_\pomdp^\States (\statesPrime) (\actions) \big\rangle
\end{align}
hold.
Since $\beh_{\States} (\states) = \beh_\StatesPrime(\statesPrime) = \statesFinal_0$, the following holds
\begin{align}
    \transitionCoalg_\pomdp^\StatesFinal \big(\beh_{\States}(\states)\big) (\actions) = \transitionCoalg_\pomdp^\StatesFinal \big(\beh_\StatesPrime(\statesPrime)\big) (\actions)
\end{align}
and therefore, component wise,
\begin{align}
    \label{eqn:conditions}
    \transitionCoalgOut_\pomdp^\States (\states) & = \transitionCoalgOut_\pomdp^\StatesPrime (\statesPrime) \nonumber \\
    \Dist (\beh_{\States}) \transitionCoalgTr_\pomdp^\States (\states) (\actions) & = \Dist (\beh_\StatesPrime) \transitionCoalgTr_\pomdp^\StatesPrime (\statesPrime) (\actions).
\end{align}
The first condition says that the observations must be equal, while the second one corresponds to equality of distributions over $\StatesFinal$, i.e. for any element $\statesFinal \in \StatesFinal$, 
the probability value $\Dist (\beh_{\States}) \transitionCoalgTr_\pomdp^\States (\states) (\actions) (\statesFinal)$ is equal to the probability value $\Dist (\beh_\StatesPrime) \transitionCoalgTr_\pomdp^\StatesPrime (\statesPrime) (\actions) (\statesFinal)$.
Using the definition of the distribution functor for discrete probability (\cref{def:distributionFunctor}), we then have that:
\begin{align}
    \sum_{\tilde{\states} \in \beh_\States^{-1}(\statesFinal)} \transitionCoalgTr_\pomdp^\States (\states) (\actions) (\tilde{\states}) = \sum_{\tilde{\states}' \in \beh_\StatesPrime^{-1}(\statesFinal)} \transitionCoalgTr_\pomdp^\StatesPrime (\statesPrime) (\actions) (\tilde{\states}').
\end{align}

Finally, we re-express the two conditions in~\cref{eqn:conditions} in a more traditional notation, obtaining:
\begin{align}
    \Dist(\observations \mid \states) & = \Dist(\observations \mid \statesPrime) & \text{(condition 1)} \nonumber \\
    \sum_{\tilde{\states} \in \beh_\States^{-1}(\statesFinal)} \Dist (\tilde{\states} \mid \states, \actions) & = \sum_{\tilde{\states}' \in \beh_\StatesPrime^{-1}(\statesFinal)} \Dist (\tilde{\states}' \mid \statesPrime, \actions) & \text{(condition 2)}
\end{align}

%
%
%
% -----------------------------------------------------------------------------------------
% -----------------------------------------------------------------------------------------
% -----------------------------------------------------------------------------------------

\section{Bayesian filtering updates}
\label{apdx:bayesianFiltering}
A belief at time $t$, a probability distribution over hidden states at time $t$, $\beliefsFiltering_{t} \coloneqq \beliefsFiltering(\states_{t})$ is defined using Bayesian filtering updates of type $\beliefUpdateFiltering : \BeliefsFiltering \times \Observations \times \Actions \to \BeliefsFiltering$, given by
    \begin{align}
        \beliefsFiltering_{t} \coloneqq & \beliefUpdateFiltering(\beliefsFiltering_{t-1}, \observations_{t}, \actions_{t-1}) \nonumber \\
        = & \prob(\states_{t} \mid \beliefsFiltering_{t-1}, \observations_{t}, \actions_{t-1}) \nonumber \\
        = & \prob(\states_{t} \mid \historiesFiltering_{t-1}, \observations_{t}, \actions_{t-1}) \big( = \prob(\states_{t} \mid \historiesFiltering_{t})\big) && \text{($\beliefsFiltering_{t-1}$ is a sufficient statistic of $\historiesFiltering_{t-1}$)} \nonumber \\
        = & \prob(\states_{t} \mid \observations_{0 \dots t-1}, \actions_{0 \dots t-2}, \observations_{t}, \actions_{t-1}) \nonumber \\
        = & \prob(\states_{t} \mid \observations_{0 \dots t}, \actions_{0 \dots t-1}) \nonumber \\
        = & \frac{\prob(\observations_{t} \mid \states_{t}, \observations_{0 \dots t-1}, \actions_{0 \dots t-1}) \prob(\states_{t}, \observations_{0 \dots t-1}, \actions_{0 \dots t-1})}{\prob(\observations_{t}, \observations_{0 \dots t-1}, \actions_{0 \dots t-1})} && \text{(Bayesian filtering)} \nonumber \\
        = & \frac{\prob(\observations_{t} \mid \states_{t}, \observations_{0 \dots t-1}, \actions_{0 \dots t-1}) \prob(\states_{t} \mid \observations_{0 \dots t-1}, \actions_{0 \dots t-1})}{\prob(\observations_{t} \mid \observations_{0 \dots t-1}, \actions_{0 \dots t-1})} \nonumber \\
        = & \frac{\prob(\observations_{t} \mid \states_{t}) \prob(\states_{t} \mid \observations_{0 \dots t-1}, \actions_{0 \dots t-1})}{\prob(\observations_{t} \mid \observations_{0 \dots t-1}, \actions_{0 \dots t-1})} && \text{(Markovianity)} \nonumber \\
        = & \frac{\prob(\observations_{t} \mid \states_{t}) \sum_{\states_{t-1}} \prob(\states_{t} \mid \states_{t-1}, \actions_{0 \dots t-1}) \prob(\states_{t-1} \mid \observations_{0 \dots t-1}, \actions_{0 \dots t-1})}{\prob(\observations_{t} \mid \observations_{0 \dots t-1}, \actions_{0 \dots t-1})} && \text{(Chapman-Kolmogorov)} \nonumber \\
        = & \frac{\prob(\observations_{t} \mid \states_{t}) \sum_{\states_{t-1}} \prob(\states_{t} \mid \states_{t-1}, \actions_{t-1}) \prob(\states_{t-1} \mid \observations_{0 \dots t-1}, \actions_{0 \dots t-2})}{\prob(\observations_{t} \mid \observations_{0 \dots t-1}, \actions_{0 \dots t-1})} && \text{(Markovianity)} \nonumber \\
        = & \frac{\prob(\observations_{t} \mid \states_{t}) \sum_{\states_{t-1}} \prob(\states_{t} \mid \states_{t-1}, \actions_{t-1}) \beliefsFiltering_{t-1}}{\prob(\observations_{t} \mid \historiesFiltering_{t-1}, \actions_{t-1})} && \text{(Definitions of $\beliefsFiltering_{t-1}$ and $\historiesFiltering_{t-1}$)} \nonumber \\
        = & \sum_{\states_{t-1}} \frac{\prob(\states_{t}, \observations_{t} \mid \states_{t-1}, \actions_{t-1}) \beliefsFiltering_{t-1}}{\prob(\observations_{t} \mid \historiesFiltering_{t-1}, \actions_{t-1})}.
    \end{align}

%Bibliography
\bibliographystyle{unsrt}  
\bibliography{AllEntriesZotero}

\end{document}